\documentclass[prd,aps,eqsecnum,floats,twocolumn]{revtex4}
\usepackage{epsfig}
\begin{document}

  \title{Optimal Strategies for Sinusoidal Signal Detection}
  
  \author{Bruce Allen} 
  \email{ballen@uwm.edu} 
  \affiliation{Department of Physics, University of
    Wisconsin - Milwaukee, P.O. Box 413, Milwaukee WI 53201}

  \author{Maria Alessandra Papa} 
  \email{papa@aei.mpg.de}  
  \author{Bernard F. Schutz} 
  \email{schutz@aei.mpg.de}
  \affiliation{Max-Planck-Institut f\"ur
    Gravitationsphysik, Albert-Einstein-Institut, Am M\"uhlenberg 1,
    D-14476 Golm, Germany}

  \date{Draft of \today}
    
  \begin{abstract}
    We derive and study optimal and nearly-optimal strategies for the
    detection of sinusoidal signals hidden in additive (Gaussian and
    non-Gaussian) noise. Such strategies are an essential part of
    algorithms for the detection of the gravitational Continuous Wave
    (CW) signals produced by pulsars.  Optimal strategies are derived
    for the case where the signal phase is not known and the product of
    the signal frequency and the observation time is non-integral.
  \end{abstract}

  \maketitle

\section{INTRODUCTION}
\label{s:intro}
A key problem in data analysis is to detect sinusoidal signals in
noise.  Such signals are often called ``lines'' or ``peaks'' because
in the Fourier domain (frequency space) they appear as spikes
(line-like features) or sharp narrow peaks in the energy spectrum of
the signal.  When the signal is large compared to the noise, such
signals are easy to identify. When they are weak, the identification
becomes more difficult.

The work in this paper was motivated by the development of algorithms
to search for Continuous Wave (CW) signals in the new generation of
interferometric gravitational-wave detectors which are either under
construction \cite{science92,physicstoday,virgo,geo600,tama300} or
planned \cite{plannedinstruments}. These signals are produced by
rapidly spinning neutron stars (pulsars).

To search for new (previously undetected) pulsars requires a search
over possible sky positions, frequencies, and pulsar spin-down
parameters. The parameter space is very large and these searches are
computationally very intensive. Moreover the searches 
will be looking for signals that are (statistically) at the lower
limit of detection sensitivity \cite{300years}.

A brute-force approach
(optimally filtering for all possible source parameters) requires
unrealistic computational resources (Petaflops), so more
sophisticated hierarchical approaches have been proposed.
When the parameter space is very large,
these approaches
retain much or all of the sensitivity of the brute-force approach but require
less computational resources. This is possible because,
in the brute-force approach, the number of grid points in parameter
space is so large that the detection threshold must be set very high
to avoid false alarms and enable confident detection.  A hierarchical
search visits fewer points in parameter space: it ignores those
below the (high) threshold that one must set in order to gain the necessary
detection confidence while examining a large parameter space. In other words 
a hierarchical search method does not ``waste'' precious computational cycles
examining regions in parameters space where, even if a signal were present,
it would not be detected confidently enough.

The hierarchical search techniques
\cite{bccs,schutz2,schutzpapa,bradycreighton} all involve a second
(so-called incoherent) stage. This stage is called ``incoherent'' because it
uses spectral rather than amplitude data. If one neglects polarization,
in all of the proposed approaches a putative signal at the second stage would
(effectively) appear in a spectrum as a sinusoidal signal at fixed frequency
and phase. The third stage of the search works only on the regions in
parameter space where significant spectral lines were identified in the second
stage.

Our paper addresses the problem of identifying these candidates, that is
``registering''
candidate sinusoidal signals. The analysis makes use of the
Neyman-Pearson criteria to identify the ``best'' statistic to use for
such identification.  In some cases, the best statistic depends upon
the expected amplitude of the signal, which is unknown.  In these
cases, we have used locally-optimal methods to identify the best
statistic in the weak-signal limit.

The analysis is complicated by several factors:
\begin{itemize}
\item The signal frequency and phase are not known in advance.
\item The signal frequency may not lie at an integer multiple of the
  Rayleigh frequency $T^{-1}$.  A signal of this type does not make an
  integer number of cycles during the observation time $T$.  We call
  such frequencies, and the corresponding signal, ``unresolved''.
\item The signal frequency must be identified with resolution less
  than $\pm (2T)^{-1}$, i.e., to within the nearest frequency bin.
\item The method must handle non-Gaussian noise in an optimal manner.
\end{itemize}
The analysis presented here addresses all of these concerns.

\section{PROBLEM DESCRIPTION AND OPTIMAL STATISTICS}
The basic problem that we consider is the following.  We are given $N$
samples of a time-domain data stream, sampled at discrete times
$t=t_j=j\Delta t$.  We denote this data by $y_j$ for
$j=0,1,\cdots,N-1$. The total observation time is $T=N\Delta t$.  The
question that we want to answer is, does the data stream $y_j$ contain
a sinusoidal signal 
\begin{equation}
\label{e:sigdef}
y_j=\epsilon {2\over N} \cos(2\pi f t_j - \phi)
\end{equation}
of constant amplitude\footnote{ The factor $2/N$ in the amplitude of
  the cosine simplifies the form of the frequency-space pdf, while
  retaining the standard definitions of the DFT.}  and frequency?  To
address this question, we make use of the theory of optimal signal
detection.  It is convenient to recast the problem in the Fourier
domain.  Denote the Discrete Fourier Transform (DFT) of the
data\footnote{This is the traditional ``physics'' definition.  The
  ``engineering'' definition has the opposite sign of $i$.}  by $x_k$:
\begin{equation}
  \label{e:fftdef}
  x_k = \sum_{j=0}^{N-1} y_j {\rm e}^{2\pi i j k/N}, \qquad
  \text{for\ } {\scriptstyle k=-N/2+1,\cdots,N/2}.
\end{equation}
Since this transformation is invertible, any question or statement
about the $y$'s can also be stated in terms of the $x$'s, hence we
will often use the term ``data'' to refer to the $x$'s rather than to
the $y$'s.  Here, and elsewhere, the symbols $x$ and $y$ without
indices refer to the collective ensemble of all the data.  For
convenience we will
assume that $N$ is a power of two.  The index $k$ will often be
referred to as a ``frequency bin''.  The frequencies that these bins
correspond to,
\begin{equation}
  \label{e:freqdef}
  f_k = {k \over N\Delta t} = {k \over T}
\end{equation}
are called ``resolved frequencies'' for reasons that will become clear
later.

In what follows, we will assume that the data $y$ is real.  In this
case, $x_k=x_{-k}^*$ where ${}^*$ denotes complex-conjugate, and both
$x_0$ and $x_{N/2}$ are real.  The data set $y$ is then exactly
equivalent to the set of $x_k$ for $k=0,\cdots,N/2$. To simplify the
mathematics, we will
assume that the average value of the $y$'s vanishes (i.e., that the DC
or average value has been removed from the data) so that $x_0=0$.  We
will also assume that there is no energy at the Nyquist frequency
$f_{N/2}$ (which in a real experiment would be enforced by appropriate
anti-aliasing filters) so that $x_{N/2}=0$. Then, the data set $y$ is
exactly equivalent to the set $x_k$ for $k=1,\cdots,N/2-1$.

We use the notation $p(x|\epsilon)$ to denote the probability
distribution function (pdf) of the data, in the presence of a signal
whose amplitude is $\epsilon$.  For example,
\begin{itemize}
\item if the (real and imaginary parts of the) noise in each frequency
  bin is independent and Gaussian with vanishing mean and unit
  variance, and the signal is a sinusoid of known phase at resolved
  frequency $f_\ell$ given by $y_j=\epsilon {2 \over N} \cos(2\pi
  f_\ell t_j-\phi)$, then
  \[
  p(x|\epsilon)= {1 \over 2\pi} {\rm e}^{-{1 \over 2} |x_\ell -
    \epsilon {\rm e}^{i\phi} |^2} \hspace{-5pt}
  \mathop{\prod_{k=1}^{N/2-1}}_{k\ne\ell} {1 \over 2\pi} {\rm e}^{-{1
      \over 2} |x_k|^2}.
  \]
  Note that since $\ell$ is an integer, and $\epsilon$ is real, the
  signal only affects the $\ell$'th frequency bin.
\item if the assumptions are the same as above, but the phase of the
  signal is unknown and uniformly distributed over the range $\phi \in
  [0,2\pi)$, then
  \[
  p(x|\epsilon)= {1 \over 2\pi} \int_0^{2\pi}\hspace{-12pt} d\phi {1
    \over 2\pi} {\rm e}^{-{1 \over 2} |x_\ell - \epsilon {\rm
      e}^{i\phi} |^2} \hspace{-5pt}
  \mathop{\prod_{k=1}^{N/2-1}}_{k\ne\ell} {1 \over 2\pi} {\rm e}^{-{1
      \over 2} |x_k|^2}.
  \]
\end{itemize}
Somewhat later, we will relax these assumptions, and give more general
forms for $p(x|\epsilon)$ where
\begin{itemize}
\item the signal frequency is not a resolved frequency,
\item the noise is not white, and
\item the noise is not Gaussian.
\end{itemize}
Note that the integration measure for $p(x|\epsilon)$ is
\[
\int dx \equiv \prod_{k=1}^{N/2-1} \int_{-\infty}^{\infty} d \Re x_k
\int_{-\infty}^{\infty} d \Im x_k,
\]
where $\Re$ and $\Im$ denote the real and imaginary parts.

The problem that we wish to solve is well-known in the theory of
signal detection.  The space of possible measurements $x_k$ for
$k=1,...,N/2-1$ is $\mathbf{R}^{N-2}$.  Our goal is to divide this
space of possible measurements into two disjoint regions $H_0$ and
$H_1$, whose union is all of $\mathbf{R}^{N-2}$.  If the observed
data lies in $H_0$ (the ``null-hypothesis region'') we will conclude
that no signal was present in the data.  If the data lies in $H_1$, we
will conclude that a signal was present. The problem we need to solve
is this: what is the best choice of $H_0$ and $H_1$?

The solution we chose is the Neyman-Pearson criterion: the best choice
is the one that gives the lowest false dismissal probability for a
given false alarm probability.  The false alarm probability $\alpha$
is the probability that a signal is detected when none is present:
\begin{equation}
\label{e:alarmdef}
\alpha=\int_{x \in H_1} dx \; p(x|0),
\end{equation}
and the false dismissal probability $\beta(\epsilon)$ is the
probability that a signal of amplitude $\epsilon$ is not found:
\begin{equation}
\label{e:dismissdef}
\beta(\epsilon) = \int_{x \in H_0} dx \; p(x|\epsilon)
\end{equation}
The Neyman-Pearson criteria leads immediately to the following rule to
partition the space of possible measurements into $H_0$ and $H_1$.
Define the likelihood function on the space of possible measurements
by
\[
\Lambda(x) = {p(x|\epsilon) \over p(x|0)}
\]
and consider the surface $\Lambda(x) = \Lambda_0 = \text{constant}$.
The Neyman-Pearson criteria leads to the following choice: take $H_0$
to be the region inside this surface, and $H_1$ to be the region
outside this surface. The value of $\Lambda_0$ that defines the
surface determines the false alarm and false dismissal probabilities.

In this paper, we will use the Neyman-Pearson criteria to define an
``optimal statistic'' which we will denote $\tau(x)$.  This is any
function of the observed data $x$ 
whose {\it level
  surfaces}
are the
same as the level surfaces of $\Lambda(x)$.  If the statistic is
greater than some threshold $\cal T$ then we conclude that a signal is
present, and if the statistic is less than the threshold $\cal T$ we
conclude that no signal was present.  The false alarm and false
dismissal probabilities are functions of this threshold $\cal T$: as
$\cal T$ is increased the false alarm probability gets smaller, and
the false dismissal probability gets larger.  In general this optimal
statistic is a function of the signal amplitude $\epsilon$.  However
we will see that for the pulsar detection problem, where $\epsilon$ is
small, the optimal statistic is effectively $\epsilon$-independent.

\section{A WORKED EXAMPLE}
\label{s:worked}
To help make these ideas concrete, we give a complete worked example,
demonstrating these ideas for the second pdf described above: a signal
of unknown phase at a resolved frequency $f_\ell$. The pdf is
\begin{equation}
\label{e:gausspdf0}
p(x|\epsilon)= {1 \over 2\pi}  \int_0^{2\pi}\hspace{-12pt} d\phi 
{1 \over 2\pi} {\rm e}^{-{1 \over 2} |x_\ell - \epsilon {\rm e}^{i\phi} |^2}
\hspace{-5pt}
\mathop{\prod_{k=1}^{N/2-1}}_{k\ne\ell} {1 \over 2\pi} {\rm e}^{-{1 \over 2} |x_k|^2}.
\end{equation}
Before continuing, it is convenient to express this in closed form.
Writing the complex data sample $x_\ell = |x_\ell| \exp(i \psi_\ell)$
in terms of its modulus $|x_\ell|$ and phase $\psi_\ell$, one has
\begin{eqnarray*}
\lefteqn{
{1 \over 2\pi}  \int_0^{2\pi}\hspace{-12pt} d\phi 
{1 \over 2\pi} {\rm e}^{-{1 \over 2} |x_\ell - \epsilon {\rm e}^{i\phi} |^2} }  \qquad & & \\
 & = & 
{1 \over 2\pi}  \int_0^{2\pi}\hspace{-12pt} d\phi 
{1 \over 2\pi}
{\rm e}^{-{1 \over 2} 
\left( |x_\ell|^2 + \epsilon^2 -2 \epsilon  \Re (x_\ell^* {\rm e}^{i \phi}) \right)
}\\
& = & 
{1 \over 2\pi} 
{\rm e}^{ -{1 \over 2} (|x_\ell|^2+\epsilon^2)} {1 \over 2 \pi} \int_0^{2\pi}
\hspace{-12pt} d\phi \;
{\rm e}^{\epsilon |x_\ell| \cos( \phi - \psi_\ell)}\cr
& = &
{1 \over 2\pi} 
{\rm e}^{ -{1 \over 2} (|x_\ell|^2+\epsilon^2)}
I_0(\epsilon |x_\ell|).
\end{eqnarray*}
The final integral has been expressed in terms of a modified Bessel
function $I_0(r)$ of the first kind
\[ 
I_0(r) = {1 \over \pi} \int_0^{\pi}\hspace{-8pt} d\theta \;{\rm e}^{r
  \cos \theta}.
\]
Thus we obtain a closed form for the pdf (\ref{e:gausspdf0})
\begin{equation}
\label{e:gausspdf}
p(x|\epsilon)= 
{\rm e}^{ -{1 \over 2} \epsilon^2}
I_0(\epsilon |x_\ell|)
\prod_{k=1}^{N/2-1} {1 \over 2\pi} {\rm e}^{-{1 \over 2} |x_k|^2}.
\end{equation}
The likelihood function is now easily found:
\begin{equation}
\Lambda(x)= {p(x|\epsilon) \over p(x|0)} =
{\rm e}^{ -{1 \over 2} \epsilon^2}
I_0(\epsilon |x_\ell|).
\end{equation}
While in a general situation, the likelihood function depends upon all
the different variables, in this particular situation it only depends
upon $|x_\ell|$.

We defined an optimal statistic $\tau$ to be any function whose level
surfaces are the same as the level surfaces of the likelihood function
$\Lambda(x)$.  In this simple situation, the likelihood function
$\Lambda(x)=\Lambda(x_1,\cdots,x_{N/2-1})$ only depends upon the
modulus $|x_\ell|$ of the amplitude in a single (the $\ell$'th)
Fourier bin.  Since it is a monotonically increasing function of
$|x_\ell|$, we can choose as an optimal statistic any monotonic
function of $|x_\ell|$, for example $|x_\ell|$ or $|x_\ell|^2$. For
historical and later convenience, let us choose as our optimal
statistic the function $\tau=|x_\ell|^2$.  This is the power in the
$\ell$'th bin.  The mean value of this statistic, the power in the
$\ell$'th bin, is
\begin{equation}
\label{e:meaninbin}
\int dx \; \tau \; p(x|\epsilon) = \int dx \; | x_\ell|^2 p(x|\epsilon) = 2 +
\epsilon^2.
\end{equation}
In the absence of a signal ($\epsilon=0$) both the real and imaginary
parts of $x_\ell$ contribute unity.

To complete the analysis of this example, we need to calculate the
false alarm and false dismissal probabilities.  We will define, for a
given value of threshold $\cal T$, the regions $H_0$ and $H_1$ by:
\begin{eqnarray*}
H_0 & = &  \left\{(x_1,\cdots,x_{N/2-1})\;\text{such that}
\;\tau=|x_\ell|^2 \le \cal T \right\}, \text{ and} \\
H_1 & = &  \left\{(x_1,\cdots,x_{N/2-1})\;\text{such that} \;
\tau=|x_\ell|^2 > \cal T \right\}. 
\end{eqnarray*}
Thus our choice of statistic gives a decision rule which has a simple
physical interpretation.  If the power in bin $\ell$ is greater than
$\cal T$, we conclude that a signal was present.  If not, we conclude
that no signal was present.

The false alarm probability (\ref{e:alarmdef}) is easy to calculate.
It is given by the following function of threshold $\cal T$.
\begin{eqnarray}
\label{e:simplealpha}
\alpha(\cal T) & = & \int_{x \in H_1} dx \; p(x|0) \cr
& = & \int_{|x_\ell|^2>\cal T} dx \; p(x|0) \cr
 & = & \int_{|x_\ell|^2>\cal T} dx 
        \prod_{k=1}^{N/2-1} {1 \over 2\pi} {\rm e}^{-{1 \over 2} |x_k|^2}\cr
 & = & 
 \mathop{\int d\Re x_\ell \int d\Im x_\ell}_{|x_\ell|^2>\cal T}
 {1 \over 2\pi} {\rm e}^{-{1 \over 2} |x_\ell|^2}\cr
 & = & 
   \int_0^{2\pi} d\psi_\ell \int_{ |x_\ell|=\sqrt{\cal T}}^\infty
   |x_\ell| d |x_\ell| 
  {1 \over 2\pi} {\rm e}^{-{1 \over 2} |x_\ell|^2}\cr
 & = & 
   \int_{ {\cal T}/2}^\infty d \left( {1 \over 2} { |x_\ell|^2 } \right)
 {\rm e}^{ - {1 \over 2} { |x_\ell|^2 }  }\cr
& = & {\rm e}^{-{\cal T}/2}
\end{eqnarray}
In this calculation, the transition from the 3rd to the 4th line is
trivial because we integrate over the all the coordinates except for
$x_\ell$. In going from the 4th to the 5th line, we have changed
variables from real and imaginary parts, to polar coordinates.

The false dismissal probability (\ref{e:dismissdef}), which depends
both upon the signal amplitude $\epsilon$ and upon the value $\cal T$
of the decision statistic threshold, is obtained with a similar
calculation.
\begin{eqnarray}
\label{e:simplebeta}
\beta(\cal T) & = & \int_{x \in H_0} dx \; p(x|\epsilon) \cr
& = & \int_{|x_\ell|^2 \le \cal T} dx \; p(x|\epsilon) \cr
& = &
{\rm e}^{ -{1 \over 2} \epsilon^2}
\int_{0}^{\sqrt{\cal T}} |x_\ell| d |x_\ell|  {\rm e}^{-{1 \over 2} |x_\ell|^2 } I_0(\epsilon|x_\ell|) \cr
& = & 
{\rm e}^{ -{1 \over 2} \epsilon^2}
\int_{0}^{{\cal T}/2} \hspace{-8pt} du \; {\rm e}^{- u } I_0(\epsilon \sqrt{2 u} )
\end{eqnarray}
This final integral can not be evaluated in closed form.  However it
is easy to check that the limit $\beta(\infty)=1$: if the threshold is
set very large, then the false dismissal probability is unity.  In a
moment, we will study the behavior of $\beta$ in the weak-signal limit
as $\epsilon \to 0$.  However, before this, it is instructive to study
the false-alarm versus false-dismissal curves for this statistic.

\begin{figure}[t]
  \psfig{file=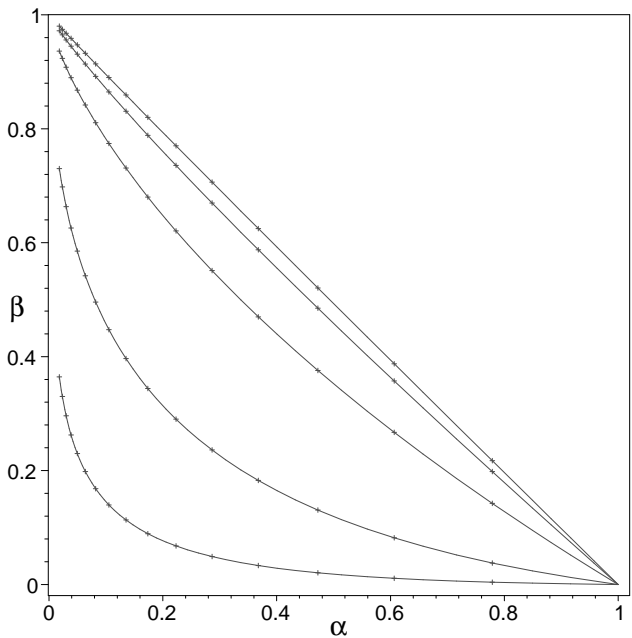,height=8.5cm}
\caption{
  The false dismissal probability $\beta({\cal T})$ as a function of
  the false alarm probability $\alpha({\cal T})$ for different values
  of the signal amplitude $\epsilon$.  The top curve has
  $\epsilon=0.2$.  Moving down, the remaining curves have
  $\epsilon=0.5,1.0,2.0,3.0$.  Along each curve, the threshold ${\cal
    T}$ varies from 0 to 8.  In the bottom right of the graph, ${\cal
    T}=0$.  The crosses mark the points where ${\cal
    T}=1/2,1,3/2,\cdots,8$. For example, with a threshold ${\cal
    T}=5.5$, if the signal amplitude is $\epsilon=3$, then the false
  alarm probability is $\alpha \approx 6.4\%$ and the false dismissal
  probability is $\beta \approx 20\%$.}
\label{f:fig1}
\end{figure}

The false alarm and false dismissal curves for this optimal detection
statistic are illustrated in Fig.~\ref{f:fig1}.  Plotting $\beta$ as a
function of $\alpha$ provides a way of describing the optimal
statistic which is completely independent of the actual choice of the
statistic.\footnote{Remember that any statistic with the same level
  surfaces as $\Lambda(x)$ is an optimal statistic.  There are an
  infinite number of different choices possible.}  However, the
relationship between the threshold $\cal T$ and the false alarm and
false dismissal probability does depend upon the choice of optimal
statistic.  Because this statistic has been chosen by the
Neyman-Pearson criterion, any other detection statistic that we choose
will have poorer performance.  Thus, for a given signal amplitude
$\epsilon$, and for a given false alarm probability $\alpha$, any
other detection statistic will have a larger false dismissal
probability $\beta$: it will lie above the illustrated curves.

\begin{figure}[t]
  \psfig{file=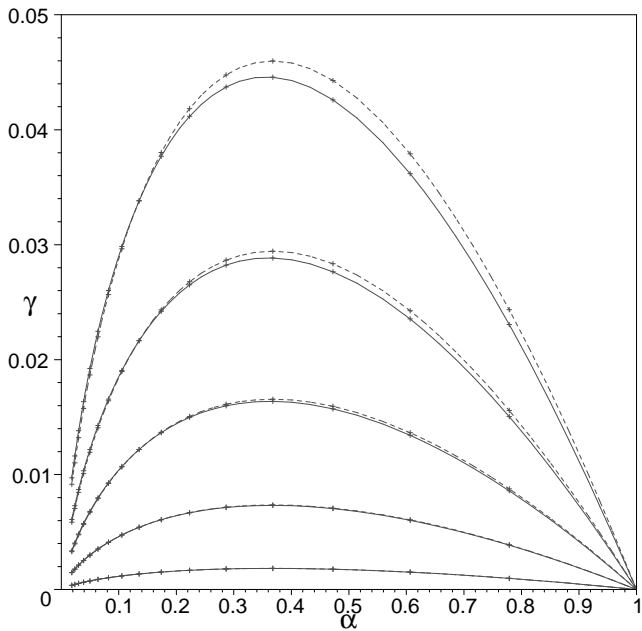,height=8.3cm}
\caption{ {\bf Solid curves}: detection probability $\gamma =
1-\alpha-\beta$ as a function of the false alarm probability $\alpha$
for different values of the signal amplitude $\epsilon=0.1,\cdots,0.5$
(moving up from the bottom curve). The crosses mark different values
of the threshold in the same way as for Fig.~\ref{f:fig1}. {\bf Dashed
curves}: the $O(\epsilon^2)$ approximation is $\gamma = \epsilon^2
{\cal T} \exp(-{\cal T}/2) /4 = -\epsilon^2 \alpha \ln \alpha /2 $.
The $O(\epsilon^4)$ approximation to $\gamma$ is not shown because on
this graph it is indistinguishable from the exact result (the solid
curves).}
\label{f:fig2}
\end{figure}

Our primary interest is in very weak signals.  For the pulsar
detection problem, we will have $\epsilon \approx 0.2$, and will be
operating on the threshold of detection where $\alpha+\beta$ is only
slightly smaller than unity. For such weak signals, it is useful to
define the quantity
\begin{equation}
 \gamma({\cal T}) = 1-\alpha({\cal T})-\beta({\cal T}).
\end{equation}
This may be considered either as a function of the threshold $\cal T$
or as a function of the false alarm probability $\alpha({\cal T})$.
This quantity $\gamma$ is the difference between the detection
probability when a signal is present, $1-\beta$, and the false alarm
probability $\alpha$.  For example, for a very weak signal, the
threshold might be set for a false alarm probability of $\alpha=15\%$.
The false dismissal probability for this weak signal might be
$\beta=84\%$.  Thus, if no signal is present, the threshold will be
exceeded $\alpha=15\%$ of the time.  If a signal is present, the
threshold will be exceeded $1-\beta=16\%$ of the time.  Roughly
speaking, the difference between these, $\gamma=1-\alpha-\beta=1\%$,
is the probability of the threshold being exceeded because the signal
was present, rather than because of the detector noise. These
weak-signal-limit curves are shown in Fig.~\ref{f:fig2}.

In the small-$\epsilon$ (weak signal) limit, it is easy to obtain an
approximate closed-form for $\beta$.  By substituting the power series
representation of the Bessel function,
\[
I_0(x) = 1 + {x^2 \over 4} + {x^4 \over 64} + {x^6 \over 2304} +
\cdots
\]
into Eqn.~(\ref{e:simplebeta}) and integrating term-by-term, one
obtains
\begin{eqnarray}
\label{e:simplegamma}
\nonumber
\lefteqn{\gamma = 1-\alpha-\beta = } & &\\
\nonumber
& & 
{\epsilon^2 \over 4} {\cal T} {\rm e}^{-{\cal T}/2}
\left[ 1 + {\epsilon^2 \over 16}({\cal T}-4) + {\epsilon^4 \over 576} 
({\cal T}^2-12{\cal T}+24) + \cdots \right] \\
\nonumber
& & = -{1\over 2}\epsilon^2  \alpha \ln \alpha \left[
 1 - {\epsilon^2 \over 8}(2+\ln \alpha)  \right. \\
 & & \left. \hspace{60pt}+ {\epsilon^4 \over 144} 
\left(  6 + 6 \ln \alpha + (\ln \alpha)^2 \right) + \cdots
\right].
\end{eqnarray}
Even at the lowest order in $\epsilon$ (the first term in square
brackets) this is a very good approximation, as shown by the dashed
curves in Fig.~\ref{f:fig2}. At the next order (the first two terms in
square brackets) the approximation is indistinguishable from the exact
result in Fig.~\ref{f:fig2} -- the solid curves.  This simplifies
matters enormously.  Although the statistics of the optimal detection
strategy depends upon the signal amplitude $\epsilon$, for small
$\epsilon$, this dependence is simple enough to be analytically
approximated.

The detection probability plays a key role in the significance of an
observation.  A hierarchical pulsar search hunts for peaks in the spectra coming from a set
of $n$ sequential time series.  For example, suppose each time series
of length $N$ is one day long.  Three months of such data would
correspond to $n=120$.  What choice of false alarm probability
$\alpha$ (or equivalently, of detection threshold $\cal T$) is
optimal?

This question is easily answered. One might guess that the best
operating point is where the detection probability $\gamma =
1-\alpha-\beta$ is maximized: in the weak signal case this is at a
threshold of ${\cal T}=2$ corresponding to a false alarm probability
$\alpha=1/e \approx 36.78\%$.  However this is not correct.  In the
absence of signal, each of the $n$ data sets is independent.  The
probability of detecting peaks in $p$ of the $n$ data sets is the same
as the probability that a coin will come up heads $p$ times in $n$
flips (if the probability of ``heads'' is the false alarm probability
$\alpha$).  This is given by the binomial distribution:
\[
\text{probability of p peaks} = \left( \begin{array}{c} n \\ p
  \end{array} \right) \alpha^p (1-\alpha)^{n-p}.
\]
Thus, in the absence of a signal, the mean number of peaks is $\alpha
n$, and its variance is $\sigma^2 = \alpha (1-\alpha) n$.  In the
presence of a signal, the mean number of peaks registered is
$(1-\beta)n$.  A good way to choose a false alarm probability (or
threshold) is to maximize the significance $s$.  This is
\begin{eqnarray}
\nonumber
s & = & { \langle \text{\# peaks} \rangle_{\rm signal}
      - \langle \text{\# peaks} \rangle_{\rm no\ signal}
       \over
        \sigma}\\
\nonumber
 & = & {(1-\beta)n - \alpha n \over \sqrt{\alpha (1-\alpha) n }} \\
\nonumber
 & = & {1-\alpha-\beta \over \sqrt{\alpha (1-\alpha)}} \sqrt{n} \\
 & = & {\gamma \over \sqrt{\alpha (1-\alpha)}} \sqrt{n}
\end{eqnarray}
The significance is easily calculated as a function of either $\alpha$
or $\cal T$.  In the weak-signal limit, it is
\[
{s \over \sqrt{n}} = {\epsilon^2 \over 4}{ {\cal T} \over \sqrt{{\rm
      e}^{{\cal T}/2} -1} } = - {\epsilon^2 \over 2} \sqrt{ \alpha
  \over 1-\alpha} \ln \alpha
\]
The significance as a function of either ${\cal T}$ or $\alpha$ has a
maximum at the threshold value ${\cal T} \approx 3.18721$
corresponding to a false alarm probability of $\alpha \approx
20.3188\%$.  The significance at this threshold/false alarm
probability is $s \approx 0.402371 \epsilon^2 \sqrt{n}$. Note that this
exhibits the expected $\sqrt{n}$ scaling in the number $n$ of spetra
analyzed. We have numerically verified that this is the optimal statistic.

\section{EXAMPLE: LOCAL PEAK DETECTION -- A NON-OPTIMAL STRATEGY}
\label{s:peakdetect}
Section~\ref{s:worked} found and analyzed the optimal (i.e.
Neyman-Pearson) peak detection strategy.  In this Section, we carry
out an identical analysis of a different (hence non-optimal) strategy.
The main purpose is to illustrate a side-by-side comparison of
different detection statistics.

We will assume that the signal and noise satisfy the same assumptions
as in Section~\ref{s:worked}, given by Eqn.~(\ref{e:gausspdf}). There,
we showed that the optimal detection strategy was to threshold on the
power $\tau=|x_\ell|^2$ in the $\ell$'th bin.  Here, we adopt a
different detection strategy.  We will say that a peak has been
detected if and only if the power $|x_\ell|^2$ in the $\ell$'th bin
exceeds the threshold $\cal T$ {\it and} is greater than the power in
either of the neighboring frequency bins.  This strategy looks for
``local peaks'' that exceed the threshold.

For this peak detection strategy, the detection region $H_1$ is
defined by
\begin{eqnarray*}
H_1 & = & \Big\{(x_1,\cdots,x_{N/2-1})\; \text{such that}\\
& & |x_\ell|^2 > {\cal T} \text{ and }
|x_\ell|^2 > |x_{\ell-1}|^2 \text{ and }
|x_\ell|^2 > |x_{\ell+1}|^2 \Big\}.
\end{eqnarray*}
In other words, the peak detection strategy is to register a peak if
the observed data set lies in $H_1$.  The null-hypothesis or no-signal
region $H_0$ is the set complement
\mbox{$H_0=\mathbf{R}^{N-2}-H_1$}: all points not lying in $H_1$.

To compare this strategy to the optimal one found in
Section~\ref{s:worked}, we calculate the false-alarm and
false-detection curves as before, and compare them with the optimal
strategy. The false alarm probability is
\begin{eqnarray*}
\label{e:localalpha}
\alpha(\cal T) & = & \int_{x \in H_1} \hspace{-18pt} dx \; p(x|0) \\
& = &
\mathop{
 \int { dx_{\ell-1} \over 2\pi }
 \int { dx_{\ell  } \over 2\pi }
 \int { dx_{\ell+1} \over 2\pi }
}_{\begin{array}{c}
|x_\ell|^2 > {\cal T} \\ 
|x_\ell|^2 > |x_{\ell-1}|^2  \\
|x_\ell|^2 > |x_{\ell+1}|^2
\end{array}}
{\rm e}^{-{1 \over 2} (|x_{\ell-1}|^2+|x_{\ell}|^2+|x_{\ell+1}|^2)}
\end{eqnarray*}
In these expressions, $\int dx_k$ denotes $\int_{-\infty}^{\infty} d
\Re x_k \int_{-\infty}^{\infty} d \Im x_k$.  Putting each of the three
integrals into polar coordinates immediately yields
\begin{eqnarray}
\label{e:localalpha2}
\nonumber \alpha(\cal T)
& = & {1 \over 2} \int_{\cal T}^\infty
\hspace{-8pt} d|x_{\ell}|^2 \; {\rm e}^{-|x_{\ell}|^2/2} \left[
{1\over 2} \int_0^{|x_{\ell}|^2} \hspace{-8pt} d|x_{\ell-1}|^2 \; {\rm
e}^{-|x_{\ell-1}|^2/2} \right]^2 \\ 
\nonumber &=& {1 \over 2}
\int_{\cal T}^\infty \hspace{-8pt} d|x_{\ell}|^2 \; {\rm e}^{ -
|x_{\ell}|^2/2} \left[\left. - {\rm
e}^{-u/2}\right\vert_{u=0}^{u=|x_{\ell}|^2}\right]^2 \\ 
\nonumber &=&
{1 \over 2} \int_{\cal T}^\infty \hspace{-8pt} d|x_{\ell}|^2 \; {\rm
e}^{- |x_{\ell}|^2/2} \left[ 1 - {\rm e}^{-|x_{\ell}|^2/2}\right]^2 \\
\nonumber & = & \int_{{\cal T}/2}^\infty \hspace{-8pt} du\; {\rm
e}^{-u} \left[ 1-{\rm e}^{-u} \right]^2 \\
& = & {1 \over 3} {\rm
e}^{-{3 \over 2} {\cal T}} - {\rm e}^{-{\cal T}}+{\rm e}^{-{1 \over 2}
{\cal T}}
\end{eqnarray}
The quantity in square brackets that appears in the intermediate steps
of this calculation is simply the probability that bins $\ell\pm 1$
contain less power than the $\ell$'th bin.  This is one minus the
false alarm probability (\ref{e:simplealpha}) of the optimal test.

As with the optimal test, the false alarm probability $\alpha({\cal
  T})$ vanishes at large threshold $\cal T \to \infty$.  However,
unlike the optimal test, the false alarm probability at zero threshold
is not unity: $\alpha({\cal T}=0)=1/3$.  This is because, even if the
threshold vanishes, to register as a peak the $\ell$'th bin must
contain more power than both adjacent bins.  When no signal is
present, this only happens $1/3$ of the time.

\begin{figure}[t]
  \psfig{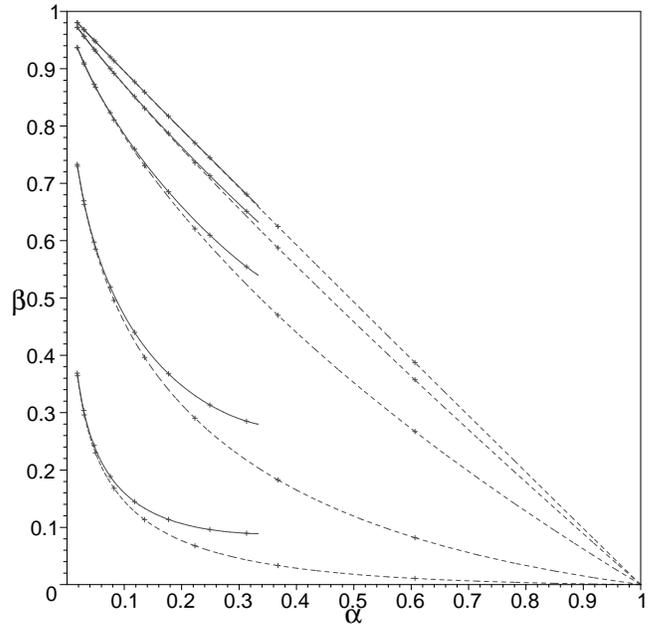}
\caption{ {\bf Solid curves}: false-dismissal $\beta$ versus
  false-alarm $\alpha$ for the non-optimal detection strategy of this
  Section.  Moving down from the top, the curves correspond to signal
  strengths $\epsilon = 0.2,0.5,1,2,3$.  Notice that the false alarm
  probability $\alpha$ is less than $1/3$ for {\it any} value of the
  threshold $\cal T$.  For comparison, the {\bf dashed curves} show
  the optimal strategy of the previous Section.  Notice that the
  optimal strategy always yields a lower false dismissal probability
  for a given false alarm probability. The crosses mark threshold
  values ${\cal T} = 1,2,\cdots,8$ increasing to the left along each
  curve.  }
\label{f:fig3}
\end{figure}

The false dismissal probability for this non-optimal peak detection
strategy can be calculated with the same methods as above.  One finds
\begin{eqnarray}
\label{e:localbeta}
\lefteqn{\beta({\cal T}) = 
 \int_{x \in H_0} \hspace{-15pt} dx \; p(x|\epsilon) } & & \\
\nonumber
& = & 1-\int_{x \in H_1} \hspace{-15pt} dx \; p(x|\epsilon)\\
\nonumber
& = & 
1 - {1 \over 2} \int_{\cal T}^\infty \hspace{-8pt} d|x_{\ell}|^2 \;
  {\rm e}^{-{1\over 2} \left( |x_{\ell}|^2 + \epsilon^2\right)}
  I_0(\epsilon |x_{\ell}|) 
 \left[1- {\rm e}^{-|x_{\ell}|^2/2}\right]^2 \\
\nonumber
& = & 1 -
{\rm e}^{ -{1 \over 2} \epsilon^2} 
\int_{{\cal T}/2}^\infty \hspace{-8pt} du \; 
I_0(\epsilon \sqrt{2 u} )\;
{\rm e}^{- u } 
\left[ 1 - {\rm e}^{- u }  \right]^2
\\
\nonumber
& = & {\rm e}^{-{1\over 4} \epsilon^2 } \hspace{-6pt} -
{\scriptstyle {1 \over 3}} {\rm e}^{- {1 \over 3} \epsilon^2 } 
 \hspace{-6pt} +
{\rm e}^{ -{1 \over 2} \epsilon^2} \hspace{-6pt}
\int_0^{{\cal T}/2} \hspace{-16pt} du \; 
I_0(\epsilon \sqrt{2 u} )\;
{\rm e}^{- u } 
\left[ 1 - {\rm e}^{- u }  \right]^2
\end{eqnarray}
As for the optimal statistic, this false dismissal probability
approaches one at large threshold ${\cal T}\to\infty$. However, unlike
the optimal test, it does not vanish at zero threshold.  Setting
${\cal T}=0$ in (\ref{e:localbeta}) on finds that
\[
\beta({\cal T}=0) = {\rm e}^{-\epsilon^2/4} - {1 \over 3} {\rm
  e}^{-\epsilon^2/3}.
\]
If the signal amplitude is small $\epsilon \to 0$ then $\beta({\cal
  T}=0)\to2/3$.  There is a $2/3$ probability of missing a small
signal at zero threshold, because one of the two neighboring frequency
bins might contain more power than bin $\ell$.

A set of false alarm/false dismissal curves for this non-optimal
statistic are shown in Fig.~\ref{f:fig3}, along with the same curves
for the optimal statistic.  Note that for a given signal strength and
false alarm probability, the false dismissal probability is always
lower for the Neyman-Pearson test. Also notice that a given value of
the threshold for one test statistic does not yield the same false
alarm probability as the same threshold value for the other statistic.
As the false alarm probability decreases, the two statistics have a
performance (false dismissal probability) that becomes increasingly
similar.  This is because at increasing values of the threshold ${\cal
  T}$, fewer and fewer peaks are rejected because the neighboring
peaks are larger.

\begin{figure}
  \psfig{file=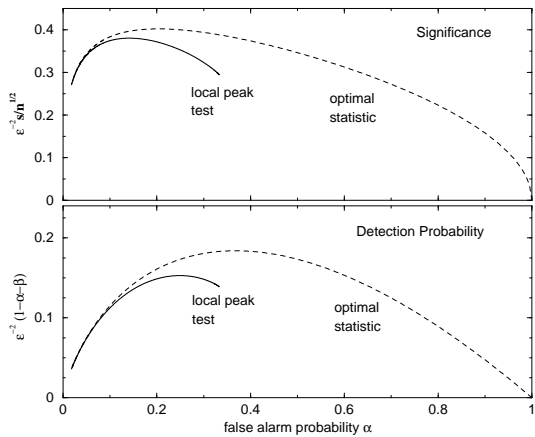,height=7cm,angle=-90}
\caption{ These graphs are a comparison of two different peak-finding
methods, in the weak signal limit (small $\epsilon$).  The dashed
curves correspond to the optimal (Neyman-Pearson) test: thresholding
on the signal power.  The solid curves correspond to the local peak
test described in this Section.  The bottom graph shows the detection
probability $\gamma/\epsilon^2=(1-\alpha-\beta)/\epsilon^2$ as a
function of false alarm probability $\alpha$.  The top graph shows the
significance $\gamma \over \epsilon^2\sqrt{\alpha(1-\alpha)}$.
Table~\ref{t:tab1} compares the properties of these curves. }
\label{f:fig4}
\end{figure}

In the small-signal limit $\epsilon\to 0$, one can use the series
expansion of the Bessel function to obtain analytic expressions for
the false alarm probability $\beta$.  The signal detection probability
is
\begin{eqnarray*}
\lefteqn{\gamma = 1-\alpha-\beta } & &\\
 & = & {\epsilon^2 {\rm e}^{-3 {\cal T}/2} \over 4}
\left[ {\cal T} 
\left(  {\rm e}^{{\cal T}} -  {\rm e}^{ {\cal T}/2} + {1 \over 3} \right)
+ \; {\rm e}^{ {\cal T}/2} - {4 \over 9} \right]  +O(\epsilon^4) \\
& = & {\epsilon^2 \over 4} \left[
\alpha {\cal T} + {\rm e}^{ - {\cal T}} -{4 \over 9} {\rm e}^{-3 {\cal T}/2}
\right] +O(\epsilon^4).
\end{eqnarray*}
This signal detection probability can't be expressed in analytic form
entirely in terms of $\alpha$ given by (\ref{e:localalpha2}).  However
we can plot it and compare with the identical curve for the optimal
strategy. This is shown in Fig.~\ref{f:fig4}, which also shows the
significance as a function of the false alarm probability.  The
comparison is shown in Table~\ref{t:tab1}.

\begin{table}
\begin{tabular}{ r|c|c }
                                                          & Optimal Test               & Local Peak Test              \\
\colrule
Maximum of $1-\alpha-\beta $ =                                    & $ 0.1839 \;\epsilon^2 $    & $ 0.1529 \;\epsilon^2 $   \\
is at threshold value $\cal T$ =                                    & $ 2.0   $                        & $ 2.0 $                        \\
and false alarm prob $\alpha$ =                               &
36.79\%                     & 24.91\%                \\
\colrule
Maximum of ${ 1-\alpha-\beta \over \sqrt{\alpha (1-\alpha)}}$ = & $0.4024 \;\epsilon^2    $ & $ 0.3806 \;\epsilon^2 $    \\
is at threshold value $\cal T$ =                                   &$ 3.187               $  & $ 3.567               $  \\
and false alarm prob $\alpha$ =                                 &  $ 20.32\%              $  &$ 14.14\%              $ 
\end{tabular}
\caption{A comparison of the optimal Neyman-Pearson detection
strategy, and the sub-optimal local peak detection method, in the
weak-signal limit. Most of these values can be read off
Fig.~\ref{f:fig4}. The top half of the table gives information about
the maximum of $\gamma = 1-\alpha-\beta$, such as the value of the
threshold at the maximum.  The bottom half of the table gives the same
information for the maximum of ${1-\alpha-\beta \over \sqrt{\alpha
(1-\alpha)}}$.  }
\label{t:tab1}
\end{table}

The primary purpose of these last two Sections was to demonstrate how
a signal detection strategy can be chosen in an optimal fashion, and
how it can be compared to a sub-optimal strategy.  In a ``real world''
situation, it may be highly desirable to apply a sub-optimal strategy,
because the mathematical model of the instrumental noise may not be
complete, and might not accurately reflect its real behavior.  In
fact, the sub-optimal method discussed in this Section has only
slightly poorer performance for the simple Gaussian noise model than
the optimal test, but may perform much better on ``real world'' data
which has correlations between different frequency bins.

In the following Section, we will apply these methods to develop
optimal tests for the case where the sinusoidal signal frequency is
not one of the exactly resolved frequencies $f_k$.

\section{COMMENTS ON THE WEAK-SIGNAL APPROXIMATION}
\label{s:weaksignal}
In the previous Sections, we studied the validity of the weak-signal
limit $\epsilon\to 0$, and made use of it when appropriate.  We will
continue to take this limit throughout the paper.  This brings up
several interesting issues.

These types of weak-signal approximations have been studied
extensively under the rubric of ``Locally optimal statistics''
\cite{kassam}.  Later in this paper, they will make treatment of
non-Gaussian noise models tractable.

In practice, the weak-signal approximation is well-justified for the
pulsar detection problem.  This is dramatically illustrated in
Fig.~\ref{f:fig2}.  This is a typical case: for $\epsilon < 1/2$ only
the lowest-order terms in $\epsilon$ need to be retained in order to
have a good approximation.  Keeping the next order terms as well gives
an extremely good approximation even for $\epsilon \approx 1$.
Typical detectable signal strengths will be $\epsilon \approx 0.2$.

In the weak-signal limit, the pdf can be well-approximated by the
first non-vanishing term in its Taylor series in $\epsilon$.  The
first derivative of $p(x|\epsilon)$ w.r.t. $\epsilon $ vanishes at
$\epsilon=0 $, because $p$ is an even function of $\epsilon$.  This is
because the phase $\phi$ of the signal is uniformly distributed in the
range $[0,2\pi)$.  The pdf is well-approximated by
\begin{equation}
\label{e:taylor1}
p(x|\epsilon) = p(x|0) + {1 \over 2} \epsilon^2 p''(x|0) + O(\epsilon^4),
\end{equation}
where ${}'$ denotes $\partial/\partial \epsilon$. The likelihood
function is then approximated by
\begin{equation}
\label{e:taylor2}
\Lambda(x) = 
{ p(x|\epsilon) \over p(x|0)}
 =1  + {1 \over 2} \epsilon^2 { p''(x|0) \over p(x|0)}.
\end{equation}
Thus in the weak signal case (neglecting second order terms in the
signal amplitude $\epsilon$) the optimal detection statistic is
independent of signal strength, and can be found from the second
derivative of the pdf at zero signal strength. This tremendously
simplifies the analysis.

The likelihood function itself, or the likelihood function minus a
constant can be used as the optimal statistic $\tau$ (for example
threshold on $\Lambda -1 $).  In the absence of signal, the mean value
of this statistic must vanish.  This follows immediately from the
definition of $\Lambda$, since
\begin{equation}
\label{e:checklam}
\int dx \; p(x|0) \; \left( \Lambda -1 \right) = 
\int dx \; \left[ p(x|\epsilon)-p(x|0) \right] = 0.
\end{equation}
In the weak signal case, keeping only terms up to a given order (say
$\epsilon^2$) in $\Lambda-1$, it is easy to show that the same
relation holds.  Hence, in the absence of a signal, the mean value of
$\Lambda(x)-1$ vanishes.  This will be useful later.

\section{OPTIMAL DETECTION OF UNRESOLVED FREQUENCY SIGNALS}
\label{s:optimalunres}
We now begin to address one of our key concerns.  The previous
Sections showed how to systematically derive and characterize a
detection strategy for the case where the weak sinusoidal signal had
unknown phase, but where, if present, the signal's frequency precisely
corresponded to one of the Fourier bins.  We now suppose that the
frequency is also a random variable, whose value is uniformly
distributed between $(f_\ell+f_{\ell-1})/2$ and
$(f_\ell+f_{\ell+1})/2$.  In other words, the signal of interest lies
somewhere between a half-bin to the left and a half-bin to the right
of the $\ell$'th frequency bin.

Before delving into the details of the analysis, it will be helpful to
briefly examine the appearance (in frequency space) of an unresolved
sinusoidal signal in the absence of noise.  Take the signal frequency
to be
\begin{equation}
\label{e:definefreq}
f_l = {l \over N \Delta t}
\end{equation}
where we do {\it not} assume that $l$ is an integer (corresponding to
one of the resolved frequencies). Let $\ell$ denote the nearest bin to
$l$, so that
\begin{equation}
\label{e:definefreq2}
l = \ell-\delta  \text{ for } \delta \in (-{1 \over 2},{1 \over 2}].
\end{equation}
Without loss of generality, we assume that the frequency $f_l$ is
between DC and Nyquist, corresponding to the range $l \in (0,N/2)$.
In the absence of noise, the signal in the time domain is given by
\[
y_j = \epsilon {2 \over N} \cos(2\pi f_l j \Delta t - \phi) = \epsilon
{2 \over N} \cos(2 \pi j l/N - \phi).
\]
Substituting this into the DFT (\ref{e:fftdef}) and using the sum of
the geometric series
\begin{equation}
\label{e:geometric}
\sum_{j=0}^{N-1} z^j = {1-z^N \over 1-z}
\end{equation}
gives Fourier amplitudes
\begin{equation}
\label{e:pattern}
x_k = \epsilon \left[ 
{\rm e}^{i\phi} D_N(k- l ) +  
{\rm e}^{-i\phi} D_N(k+ l ) \right],
\end{equation}
where the function $D_N$ is the {\it Dirichlet Kernel}:
\begin{equation}
\label{e:DNdef}
D_N(z) = {\rm e}^{i \pi z \left(1 - {1 \over N} \right)} {\sin (\pi z)
  \over N \sin (\pi z/N)}
\end{equation}
As described following equation (\ref{e:freqdef}), the range of the
frequency index $k$ is $1,\cdots,N/2-1$.  Since $D_N(z)$ vanishes for
all integer arguments except for zero, where its value is $D_N(0)=1$,
in the resolved-frequency case where $ l $ is an integer, one has $x_k
= 0$ for $k \ne l $, and $x_ l =\epsilon {\rm e}^{i\phi}$. In the
unresolved case, the signal energy is not confined to the $\ell$'th
bin, and forms a characteristic pattern of ``side-lobes'' in the
nearby frequency bins.

If the signal frequency is unresolved ($l$ non-integer)) 
the optimal statistical test will not only involve
data from the $\ell$'th bin. The adjacent frequency bins also contain
part of the signal energy, and we will shortly find that the
statistically optimal search also takes into account their content (in
the sense of energy and information).

One can simplify the form of the Dirichlet kernel with several
approximations\footnote{Further justification for these approximations
may be found in Section~\ref{s:window} and Fig.~\ref{f:fig7}.}. Our
primary interest is to extract as much useful information as possible
from the Fourier amplitudes in the bins near bin $\ell$.  Because
$D_N(z)$ is strongly peaked at $z=0$ and falls off $\sim z^{-1}$ away
from it, one may neglect the second term in (\ref{e:pattern}) and
concentrate on the first term. In addition, in practical applications,
$N$ will be large enough (greater than $10^5$) that the $1/N$ term in
the exponential of $D_N$ can be neglected.  Finally, since we will be
interested in the Fourier amplitudes in nearby bins, $|z|<<N$, which
means that the denominator $N \sin (\pi z/N)$ is well-approximated by
$\pi z$. This leaves us with
\[
x_k \approx \epsilon {\rm e}^{i \phi} \omega( k-l ),
\]
where the coefficients
\begin{eqnarray}
\label{e:definew}
\nonumber
w( z ) & = & 
{\rm e}^{i\pi z}
{\sin \pi z \over \pi z} \\
\nonumber
& = &
{\rm e}^{i\pi z} j_0(\pi z ) \\
& = &
{\rm e}^{i\pi z} {\rm sinc}( z ).
\end{eqnarray}
Here $j_0$ is a spherical Bessel function, and we have used Woodward
and Bracewell's definition of the sampling function $\rm sinc$.

We now suppose that the signal of interest is distributed, with equal
probability, anywhere between $\pm 1/2$ a frequency bin from the
$\ell$'th bin, and write an expression for the pdf of the data.  If,
as before, the signal phase $\phi$ is a uniformly distributed random
variable, and if the instrument noise is Gaussian and satisfies the
same assumptions as before, one has $p(x|\epsilon)= $
\begin{equation}
\label{e:defineprob}
{1 \over 2\pi} \int_{-1/2}^{1/2} \hspace{-12pt} d\delta \;  \int_0^{2\pi}\hspace{-12pt} d\phi 
\prod_{k=1-\ell}^{N/2-1-\ell}
{1 \over 2\pi} {\rm e}^{-{1 \over 2} |x_{k+\ell} - 
\epsilon  \omega(k+\delta) {\rm e}^{i\phi} |^2}.
\end{equation}
In this expression, which involves a product over all frequency bins,
the index $k$ has been shifted so that $k=0$ labels the $\ell$'th bin.

When searching for a signal peak in the vicinity of the $\ell$'th bin,
there are practical reasons (computational efficiency and algorithm
structure) why it is desirable to use only information from (some
small number of) nearby bins\footnote{Section~\ref{s:window} and
Fig.~\ref{f:fig7} show that virtually all the information is within a
few bins from the $\ell$'th bin.}.  Fortunately for us, the
Neyman-Pearson criteria can be easily derived for this more limited
information: we merely write down the pdf for the part of the data
(the nearby bins) which are available to us.  From this point on, we
will assume that our search for a signal in the vicinity of the
$\ell$'th frequency bin is restricted to $2P+1$ bins.  These are the
$\ell$'th bin itself, and $P$ frequency bins to its left and to its
right.  For this restricted data set, the pdf is $p(x|\epsilon)= $
\begin{equation}
\label{e:restrictedpdf}
{1 \over 2\pi} \int_{-1/2}^{1/2} \hspace{-12pt} d\delta
\; \int_0^{2\pi}\hspace{-12pt} d\phi \prod_{k=-P}^{P} {1 \over 2\pi}
{\rm e}^{-{1 \over 2} |x_{k+\ell} - \epsilon 
 \omega(k+\delta) {\rm e}^{i\phi} |^2}.
\end{equation}
One may now easily write down the likelihood function, and an optimal
statistic, in the weak signal limit, making use of
Eqn.~(\ref{e:taylor1}) and (\ref{e:taylor2}). It is easily verified
that there are no terms of order $\epsilon$.  Writing the pdf in the
form
\begin{equation}
p(x|\epsilon)= {1 \over 2\pi} \int_{-1/2}^{1/2} \hspace{-12pt} d\delta
\; \int_0^{2\pi}\hspace{-12pt} d\phi \; {\rm e}^{W(\epsilon)}
\end{equation}
where
\[
W(\epsilon) \equiv \sum_{k=-P}^{P} \left\{ -{1 \over 2} \left|
    x_{k+\ell} - \epsilon \omega(k+\delta) {\rm e}^{i\phi} \right|^2 -
  \ln 2\pi \right\},
\]
and taking two derivatives w.r.t. $\epsilon$, one has
\begin{eqnarray}
\label{e:uglyform}
\nonumber
p''(x|0) & = &{1 \over 2\pi} \int_{-1/2}^{1/2} \hspace{-12pt} d\delta
\; \int_0^{2\pi}\hspace{-12pt} d\phi\;  
{\rm e}^{W(0)}
\left[ \left(W'(0) \right)^2 + W''(0) \right] \\
& = &  p(x|0) \int_{-1/2}^{1/2} \hspace{-12pt} d\delta
 \int_0^{2\pi}\hspace{-4pt} {d\phi \over 2\pi} \;
\left[ \left(W'(0) \right)^2 + W''(0) \right].
\end{eqnarray}
We will do similar calculations later, in
much less detail.  The derivatives are easily evaluated:
\begin{eqnarray}
W'(0) & = & \left. { dW \over d\epsilon} \right|_{\epsilon=0} =
\sum_{k=-P}^{P} \Re \left( x^*_{k+\ell} \omega(k+\delta) 
{\rm e}^{i\phi} \right), \\
W''(0) & = &  \left. { d^2W \over d\epsilon^2} \right|_{\epsilon=0} =
- \sum_{k=-P}^{P}| \omega(k+\delta)|^2.
\end{eqnarray}
The integral $d\phi$ of $W'(0)^2$ is evaluated by noting that for any
complex numbers $A$ and $B$
\begin{eqnarray}
  \label{e:intdphi}
  \nonumber
  \lefteqn{
    \int_0^{2\pi}\hspace{-4pt} {d\phi \over 2\pi}
    \Re \left(A {\rm e}^{i\phi} \right)
    \Re \left(B {\rm e}^{i\phi}\right)} & & \\
  \nonumber
  & = &|A||B| \int_0^{2\pi}\hspace{-4pt} {d\phi \over 2\pi}
  \cos(\phi-\psi_A) \cos(\phi-\psi_B) \\
  \nonumber
  & = & 
  {|A||B| \over 2}  \int_0^{2\pi}\hspace{-4pt} {d\phi \over 2\pi}
  \left[ \cos(\psi_B - \psi_A)+\cos(2\phi-\psi_A-\psi_B) \right] \\
  \nonumber
  & = & {1\over 2} |A||B| \cos(\psi_B - \psi_A) \\
  & = & {1\over 2} \Re(AB^*)
\end{eqnarray}
Making use of this, the inner integral in (\ref{e:uglyform}) gives
\begin{eqnarray*}
\lefteqn{ \int_0^{2\pi}\hspace{-4pt} {d\phi \over 2\pi} \;
\left[  \left(W'(0) \right)^2 + W''(0) \right]} & & \\
& = &{1 \over 2} \Re \sum_{r=-P}^P  \sum_{r'=-P}^P 
 x^*_{r+\ell} x_{r'+\ell}  \omega(r+\delta) \omega^*(r'+\delta) \\
& & -  \sum_{r=-P}^P |  \omega(r+\delta)|^2.
\end{eqnarray*}
Substituting this back into expression (\ref{e:uglyform}) for the
second derivative of the pdf yields
\begin{equation}
\label{e:niceform}
{ p''(x|0) \over p(x|0)} =
{1 \over 2} \sum_{r,r'=-P}^P  x^*_{r+\ell} M_{rr'}  x_{r'+\ell} - \sum_{r=-P}^P M_{rr}.
\end{equation}
Here, $M_{rr'}$ is a $(2P+1)$-dimensional square, symmetric, real,
positive-definite matrix.  Making use of the definition of $\omega$ in
Eqn.~(\ref{e:definew}) gives
\begin{eqnarray}
\label{e:definem}
{\bf M} & = & M_{rr'} =   \int_{-1/2}^{1/2} \hspace{-12pt} d\delta \;
 \omega(r+\delta)  \omega^*(r'+\delta) \\
\nonumber
& = & 
 (-1)^{r-r'}  \int_{-1/2}^{1/2} \hspace{-12pt} d\delta \;
 j_0(\pi(r+\delta)) j_0(\pi(r'+\delta)).
\end{eqnarray}
Adopting the Einstein summation convention (the repeated indices $r$
and $r'$ are summed from $-P$ to $P$) and substituting
(\ref{e:niceform}) into the weak-signal approximation
(\ref{e:taylor2}) of the likelihood function one obtains
\begin{eqnarray}
\label{e:optimaluniform}
\Lambda(x) -1 & = &
 {\epsilon^2 \over 2} 
\left( {1 \over 2} x^*_{r+\ell} M_{rr'}  x_{r'+\ell} - M_{rr}\right) \cr
 & = & {\epsilon^2 \over 2} 
  \left( {1 \over 2} x^*_{r+\ell} x_{r'+\ell} -  \delta_{rr'}\right) M_{rr'}.
\end{eqnarray}
In the absence of a signal, Eqn.~(\ref{e:checklam}) shows that the
mean value of $\Lambda-1$ must vanish.  This is clearly the case,
since under our assumptions, in the absence of a signal, the mean
value of $ x^*_{r+\ell} x_{r'+\ell}$ is $2 \delta_{rr'}$, where $
\delta_{rr'}$ is the Kronecker Delta.

We note that the formalism of this Section can be trivially adapted to
the case where the frequency of the signal lies in any desired range
$\pm \Delta$ around the $\ell$'th bin.  The only change is that in
Eqn.~(\ref{e:definem}) one makes the transformation
\begin{equation}
\label{e:arbband}
\int_{-1/2}^{1/2} d\delta \; \rightarrow {1 \over 2 \Delta}
\int_{-\Delta}^{\Delta }  d\delta .
\end{equation}
In the limit $\Delta \to 0$, it's easy to see that $M_{00}=1$ and all
other components of $M_{rr'}=0$. The results are then identical to the
resolved-frequency case of Section~\ref{s:worked}.

The results of this Section can be summarized in a few lines.  In
Section~\ref{s:worked} we studied the case where the signal frequency
was exactly resolved.  In this case, we found that the optimal
statistic was the power in that bin.  Thresholding on this statistic
gave the lowest false dismissal probability for a given false alarm
probability.  In this Section, after assuming that the signal
frequency is uniformly distributed around bin $\ell$, we have found
that the optimal statistic (in the weak-signal case) is to threshold
on the bilinear quantity (\ref{e:optimaluniform}).  We can choose
(from the value of $P$) how much of the data around the given bin to
use. If $P=0$ we recover the power statistic of
Section~\ref{s:worked}.  If $P$ is larger, then additional information
from neighboring bins also gets added, and the test performs better.
In the following Sections, we will analyze the performance of this
test, using the methods of Section~\ref{s:peakdetect} to compare the
optimal statistics for different values of $P$.

\section{PROPERTIES OF THE MATRIX M}
Let us begin by exhibiting the $(2P+1)$-dimensional matrix ${\bf M}$,
given by Eqn.~(\ref{e:definem}). It's easy to integrate
(\ref{e:definem}) to get an exact expression for the matrix in terms
of sine- and cosine-integral functions $\rm Si$ and $\rm Ci$.  On the
diagonal (no summation convention on $n$)
\[
M_{nn} = {4 \over \pi^2 (4n^2-1)} + {1 \over \pi}{\rm Si} (\pi(2n+1))
- {1 \over \pi}{\rm Si} (\pi(2n-1)),
\]
and off the diagonal
\[
M_{mn} ={ {\rm C}(2m-1)-{\rm C}(2m+1)-{\rm C}(2n-1)+{\rm C}(2n+1)
  \over 2\pi^2(n-m)},
\]
where ${\rm C}(x) \equiv {\rm Ci}(\pi x) - \ln x$. In these equations,
the range of the subscripts $n,m$ is $-P,\cdots,P$.

The ``central'' element of $\bf M$ has row and column number zero.
The matrix extends away from this central element by an amount
determined by the value of $P$.  For example, if $P=2$ one has the
$5$-dimensional matrix:
\[
{\bf M}= 0.7737 {\scriptsize \tt \left[
\begin{array}{rrrrr}
 0.0181 &  0.0422 & {\bf -0.0169} & -0.0366 & -0.0169 \\ 
 0.0422 &  0.1017 & {\bf -0.0761} & -0.0761 & -0.0366 \\ 
{\bf -0.0169} & {\bf -0.0761} & {\bf  1.0000} & {\bf -0.0761} & {\bf -0.0169} \\ 
-0.0366 & -0.0761 & {\bf -0.0761} &  0.1017 &  0.0422 \\ 
-0.0169 & -0.0366 & {\bf -0.0169} &  0.0422 &  0.0181
\end{array}
\right]},
\]
where the $0$'th row and column are highlighted, and we have taken out
an overall factor of $M_{00}$.  Note that this matrix is invariant
under reflection about both diagonals, so it can be presented by
listing just the $P+1$-dimensional block of elements with non-negative
row and column number.

Because the matrix $\bf M$ is real and symmetric, it can be
diagonalized by a similarity transformation
\begin{equation}
{\bf D} = {\bf O}^{-1} {\bf M O}
\end{equation}
where $\bf O$ is an orthogonal square matrix ${\bf O}^t = {\bf
  O}^{-1}$, and $\bf D$ is diagonal.  Because $\bf M$ is positive, its
eigenvalues are all real and positive.  To six decimal places of
accuracy, for the first few values of $P$, the eigenvalues of $\bf M$
are given by:
\begin{equation}
\label{e:lam0num}
\lambda_0 = 7.73695 \times 10^{-1} \text{   for } P=0,
\end{equation}
\begin{eqnarray}
\lambda_0 & = &  7.82774 \times 10^{-1} \cr
\lambda_1 & = &  1.37549 \times 10^{-1}\cr
\lambda_2 & = &  1.07687 \times 10^{-2} \text{   for } P=1,
\end{eqnarray}
\begin{eqnarray}
\lambda_0 & = & 7.83230 \times 10^{-1}  \cr
\lambda_1 & = & 1.64608 \times 10^{-1}  \cr
\lambda_2 & = & 1.12358 \times 10^{-2}\cr
\lambda_3 & = & 8.16859 \times 10^{-5} \cr
\lambda_4 & = & 1.53779 \times 10^{-6 } 
 \text{   for } P=2,
\end{eqnarray}
\begin{eqnarray}
\lambda_0 & = & 7.83317 \times 10^{-1} \cr
\lambda_1 & = & 1.76172 \times 10^{-1} \cr
\lambda_2 & = & 1.13227 \times 10^{-2} \cr
\lambda_3 & = & 1.20531 \times 10^{-4} \cr
\lambda_4 & = & 1.91042 \times 10^{-6} \cr
\lambda_5 & = & 3.03979 \times 10^{-9 }\cr
\lambda_6 & = & 2.72000 \times 10^{-11}
 \text{   for } P=3.
\end{eqnarray}
We will see shortly that these eigenvalues determine the false alarm
and false dismissal probabilities for the corresponding threshold
statistics/tests.

The case analyzed in Section~\ref{s:worked}, where the signal
frequency is resolved, and a one-point test is used, corresponds to
setting $P=0$ and having $\lambda_0=1$.  This is the limit when the
frequency band (\ref{e:arbband}) over which the signal is distributed
is very small, and centered around a bin frequency.  In the opposite
limit where the frequency band $\pm \Delta$ is large, the matrix $\bf
M$ approaches something proportional to the identity matrix, with a
large number of nearly-equal eigenvalues.

\section{PERFORMANCE OF THE OPTIMAL TEST FOR UNRESOLVED SIGNALS}
The situation we are considering is defined by the pdf given in
Eqn.~(\ref{e:defineprob}). We will suppose that we have implemented a
search for sinusoidal signals (in the weak signal limit) using the
thresholding statistic defined by Eqn.~(\ref{e:optimaluniform}), for a
particular value of $P$.  We will call such a test the ``$2P+1$-point
test''. For example, the ``five point test'' makes use of the data
samples in the five bins nearest to some central bin, to determine if
a sinusoidal signal is present in that central bin.

Our goal is to determine the false-alarm and false-dismissal curves
for different values of $P$.  In this way, one can quantify the loss
of performance that arises from throwing away the additional
information coming from bins located away from the bin of interest.

Let us first calculate the false-alarm probability for the
$2P+1$-point test.  This is easy because it only involves the
probability distribution $p(x|0)$ (and its second derivative)
for vanishing signal strength, which
is an independent Gaussian in each frequency bin.  We choose, as our
optimal statistic, the quantity
\begin{equation}
\label{e:defineopt}
{\tau} \equiv {\bf x}^\dagger {\bf M} {\bf x}
\end{equation}
where ${\bf x}$ is a vector of (frequency space) data around the bin of
interest.  This differs from $\Lambda-1$ by a data-independent
constant term, $\epsilon^2\over 2$, so it has the same level surfaces. 
Thus, for the
3-point test, the optimal statistic to threshold on would be
\[
{\tau}= { [x^*_{\ell-1} \quad x^*_{\ell} \quad
  x^*_{\ell+1}] \left[
{\scriptsize \tt 
   \begin{array}{rrr}
   0.0787 & -0.0589 & -0.0589 \\ 
  -0.0589 &  0.7737 & -0.0589 \\ 
  -0.0589 & -0.0589 &  0.0787
  \end{array}
}
\right] \left[
  \begin{array}{c}
  x_{\ell-1} \cr
  x_{\ell}\cr
  x_{\ell+1}
  \end{array}
\right]}.
\]
In the absence of signal, each of the $x_j$ is an independent random
Gaussian variable with zero mean and unit variance.  Thus, if $U$ is a
unitary matrix, the column vector of variables ${\bf U x}$ are also
independent random Gaussian variables with zero mean and unit
variance. Since the orthogonal matrix ${\bf U = O}^{-1}$ that
diagonalizes $\bf M$ is unitary, the statistical properties of the
optimal statistic ${\tau}$ (\ref{e:defineopt}) are the same as those
of a random variable
\[ 
{\tau}= \sum_{r=0}^{2P} \lambda_r |z_r|^2
\]
where each $z_r$ is an independent variable whose real and imaginary
parts have independent Gaussian pdfs with zero mean and unit variance.
Note that the pdf of $u= |z_r|^2$ is exponential with mean and variance 
equal to $2P$.

The pdf of the statistic ${\tau}$ is easily computed using generating
functions.  Suppose that ${\tau}$ is any random variable, and
$p({\tau}) d{\tau}$ is its probability density. We define the
generating function $\bar p(\xi)$ to be the expected value of ${\rm
  e}^{i \xi {\tau}}$
\[
\bar p(\xi) = \int_{-\infty}^\infty d{\tau} p({\tau}) {\rm e}^{i \xi
  {\tau}}.
\]
This is basically the Fourier transform of the pdf. It makes it simple
to compute the pdf of a random variable that is a sum of other random
variables.  Since
\[
{\tau} = \sum_{r=0}^{2P} \lambda_r u_r
\]
where each $u_r$ is a real random variable with pdf
\[
p(u) du = \cases{ 0 & for $u<0$ \cr {1 \over 2} {\rm e}^{-u/2} du &
  for $u \ge 0$ }
\]
the generating function for the pdf of ${\tau}$ (in the absence of a
signal) is
\begin{eqnarray}
\bar p(\xi)  & =&   \prod_{r=0}^{2P}\left[ \int_0^\infty 
\hspace{-8pt}
du_r \;
{1 \over 2} {\rm e}^{-u_r/2}\right] {\rm e}^{i \xi {\tau}} \cr
& =&   \prod_{r=0}^{2P}\left[ \int_0^\infty 
\hspace{-8pt}
du_r \;
{1 \over 2} {\rm e}^{-u_r/2}\right] {\rm e}^{i \xi (\lambda_0 u_0 +
\cdots \lambda_{2P} u_{2P}) } \cr
& = &
\prod_{r=0}^{2P}{1 \over 2} \int_0^\infty 
\hspace{-8pt}
du_r \;
{\rm e}^{( i \xi \lambda_r -1/2)u_r}\cr
& = & \prod_{r=0}^{2P} (1- 2 i \xi \lambda_r)^{-1}
\end{eqnarray}
This closed form for the generating function $\bar p$ makes it
possible to find the probability distribution of the optimal statistic
${\tau}$ in the absence of a signal.

To determine $p$ from $\bar p$, we invert the Fourier transform
\[
p({\tau}) = {1 \over 2\pi} \int_{-\infty}^\infty d\xi \; \bar p(\xi)
{\rm e}^{-i \xi {\tau}}.
\]
This gives
\begin{eqnarray}
p({\tau}) & = & {1 \over 2\pi} \int_{-\infty}^\infty d\xi \; {\rm e}^{-i \xi
{\tau}}  \prod_{r=0}^{2P} {i  \over 2 \lambda_r }\left( \xi +  {i
 \over 2 \lambda_r } \right)^{-1}.
\end{eqnarray}
The integral clearly vanishes for ${\tau}<0$, because the integrand
has all of its poles in the complex $\xi$-plane below the real-$\xi$
axis.  If ${\tau}<0$, the sign of the exponential term permits the
contour of integration to be closed in the upper-half $\xi$-plane.
Since there are then no poles contained inside the integration path,
Cauchy's theorem implies that $p({\tau})=0$ for ${\tau}<0$.

To find a closed form for $p({\tau})$ when ${\tau}>0$, one must close
the integration contour in the lower-half $\xi$-plane.  The residue
theorem then implies that $p({\tau})$ is a sum over the resides of the
poles, which are located at $\xi = - i/2\lambda_r$. One obtains
\begin{eqnarray}
\label{e:poftau}
\nonumber
p({\tau}) & = & \sum_{r=0}^{2P} \left[ {{\rm e}^{-{\tau}/2\lambda_r} \over 2
\lambda_r} \mathop{ \prod_{r'=0}^{2P}}_{r'\ne r} \left( 1- {
\lambda_{r'} \over \lambda_r} \right)^{-1} \right]\\
& = &
\sum_{r=0}^{2P} {c_r \over 2 \lambda_r} \; {\rm e}^{-{\tau}/2\lambda_r} .
\end{eqnarray}
Here, we have introduced the set of $2P+1$ weights $c_0,\cdots,
c_{2P}$ defined by
\[
c_r \equiv \mathop{ \prod_{r'=0}^{2P}}_{r'\ne r} \left( 1- {
    \lambda_{r'} \over \lambda_r} \right)^{-1}.
\]
(Note: if $P=0$ then $c_0=1$).  These weights have several interesting
properties.  In particular
\begin{eqnarray}
\label{e:wprops}
\sum_{r=0}^{2P} c_r & = & 1, \text{  and  }\\
 \sum_{r=0}^{2P} c_r \lambda_r
& = & \sum_{r=0}^{2P} \lambda_r = M_{rr}.
\end{eqnarray}
These weights simplify the notation in what follows.

The false alarm probability $\alpha({\cal T})$ can now be obtained by
straightforward integration:
\begin{eqnarray}
\label{e:alphaform}
\nonumber
\alpha({\cal T}) & = & \int_{\cal T}^\infty d{\tau} \; p({\tau}) \\
 & = & \sum_{r=0}^{2P} c_r   {\rm e}^{-{\cal T} /2\lambda_r}.
\end{eqnarray}
It follows from (\ref{e:wprops}) that $\alpha(0)=1$.

Our calculations assume that the eigenvalues $\lambda_r$ are distinct
(as is the case here).  If $m$ of them were equal then a polynomial of
order $m-1$ in $\tau$ would appear on the r.h.s. of (\ref{e:poftau})
and a polynomial of order $m-1$ in ${\cal T}$ would appear on the
r.h.s. of (\ref{e:alphaform}).

For concreteness, we give the numerical form of the false alarm
functions for the first few values of $P$.  The subscript on $\alpha$
denotes $2P+1$: the number of points used in the test.
\begin{eqnarray*}
\alpha_1({\cal T}) & = & 
{\rm e}^{- 0.646249\,{\cal T}}\\
\alpha_3({\cal T}) & = &
{{\rm e}^{- 0.63875\,{\cal T}+0.207097}}-
{{\rm e}^{-3.6351\,{\cal T} -1.46410}} \\
& & + \,
{{\rm e}^{- 46.430\,{\cal T} -6.73815}}\\
\alpha_5({\cal T}) & = & 
{{\rm e}^{- 0.63840\,{\cal T} + 0.250487}}- 
{{\rm e}^{- 3.0375\,{\cal T} - 1.25272}} \\
& & + \, 
{{\rm e}^{- 44.500\,{\cal T} -6.83620}}-
{{\rm e}^{- 6121.0\,{\cal T}-21.6738}} \\
& &
+\,{{\rm e}^{- 325140.0\,{\cal T}-37.5716}}
\end{eqnarray*}
The false dismissal probability $\beta$ is a bit more challenging to
calculate. However for the weak-signal case of interest, it is still
possible.

To find false dismissal probability $\beta$ we begin by writing the
pdf for the weak signal case as
\begin{eqnarray*}
p(x|\epsilon) &=& p(x|0) + {1 \over 2} \epsilon^2 p''(x|0)  \\
&=& p(x|0)\left( 1 + {1 \over 2} \epsilon^2 { p''(x|0)
\over p(x|0) }
\right) \\
& = & p(x|0) \left( 1 + {1 \over 2} \epsilon^2 \left( {1 \over 2}x^*_{r+\ell}
M_{rr'}  x_{r'+\ell} -  M_{rr}\right) \right)\\
& = & p(x|0) \left( 1 + {1 \over 2} \epsilon^2 
\left({{\tau} \over 2} - M_{rr}\right) \right)
\end{eqnarray*}
where ${\tau}$ is the optimal statistic (\ref{e:defineopt}).  From
this, we can immediately write an expression for the generating
function of $p({\tau}|\epsilon)$ to lowest order in $\epsilon$,
\[
\bar p(\xi|\epsilon) = \prod_{r=0}^{2P} \left[ \int_0^\infty
  \hspace{-10pt} du_r \; {1 \over 2} {\rm e}^{-u_r/2}\right] {\rm
  e}^{i \xi {\tau}} \left( 1 + {1 \over 2} \epsilon^2 \left({{\tau}
      \over 2} - M_{rr}\right) \right),
\]
where as before ${\tau}=\lambda_0 u_0 + \cdots \lambda_{2P} u_{2P}$.
Since differentiating w.r.t. $\xi$ brings down a factor of $i{\tau}$,
one has
\begin{equation}
\bar p(\xi|\epsilon) =
 \left[ 1 + {1\over 2} \epsilon^2 \left({1 \over 2 i}{d \over d\xi} - 
 M_{rr}\right) \right] p(\xi|0).
\end{equation}
This relation is easily inverted to find a lowest-order formula for
$p({\tau}|\epsilon)$.  We simply integrate the new term by parts:
\begin{eqnarray*}
 & & {1 \over 2\pi} \int_{-\infty}^\infty d\xi\; {\rm e}^{-i \xi {\tau}} 
{d \bar p \over d\xi} \\
& = &  {1 \over 2\pi} \int_{-\infty}^\infty
{d \over d\xi} \Big[
 {\rm e}^{-i \xi {\tau}}  \bar p(\xi) \Big] + i {\tau}  {\rm e}^{-i \xi {\tau}}
 \bar p(\xi) \; d\xi \\
& = & {\tau} p({\tau}) = {\tau} p({\tau}|0).
\end{eqnarray*}
Thus we find a formula for the pdf of the optimal statistic ${\tau}$
in the small-$\epsilon$ limit:
\[
p({\tau}|\epsilon) = p({\tau}|0) \left( 1 + {1 \over 2} \epsilon^2
  \left({ {\tau} \over 2} - M_{rr}\right) \right).
\]
Since the pdfs on both sides are normalized, an important consequence
of this is that the mean value of the test statistic in the absence of
a signal is
\[
M_{rr} = \sum_{r=0}^{2P} \lambda_r = {1 \over 2} \int_0^\infty
d{\tau}\; {\tau}\, p({\tau}).
\]
This is because the mean value of the likelihood function in the
absence of a signal is unity.  It's also easy to show that
$\int_0^\infty \alpha({\cal T}) \,d{\cal T} = 2M_{rr}$.

\begin{figure}[t]
  \psfig{file=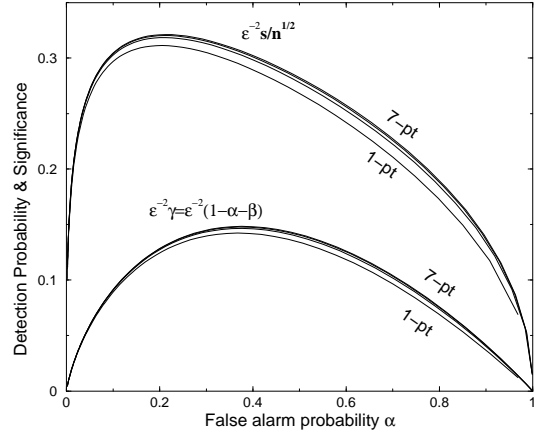,height=7cm,angle=-90}
\caption{ {\bf Bottom four curves:} The detection probability
  $\epsilon^{-2} \gamma = \epsilon^{-2}(1-\alpha-\beta)$ is plotted as
  a function of the false alarm probability $\alpha$, for the 1,3,5,
  and 7-point optimal tests defined by Eqn.~(\ref{e:defineopt}), in
  the weak-signal limit.  While using the additional information in
  the neighboring bins does improve the detection probability, the
  improvement is slight.  {\bf Top four curves:} The significance
  $\epsilon^{-2} s/\sqrt{n}$ is plotted for the same 1,3,5, and
  7-point tests, in the weak-signal limit. The maxima of the eight
  curves is given in Table~\ref{t:tab2}.}
\label{f:fig5}
\end{figure}

From this it is straightforward to calculate the false dismissal
probability
\begin{eqnarray*}
\beta({\cal T}) & = & \int_0^{\cal T} d{\tau} \; p({\tau}|\epsilon) \\
& = &  \left( 1 - {\epsilon^2 \over 2}  M_{rr}\right) \int_0^{\cal T} d{\tau} \; p({\tau}|0)
 + {1 \over 4} \epsilon^2 \int_0^{\cal T} d{\tau} \; {\tau} \, p({\tau}|0) \\
& = & \left( 1 - {\epsilon^2 \over 2}  M_{rr}\right)(1-\alpha({\cal T}))\\
& & 
+ {\epsilon^2 \over 4}
 \sum_{r=0}^{2P} c_r \left(2\lambda_r -({\cal T}+2\lambda_r)  {\rm e}^{-{\cal T}/2\lambda_r} \right)\\
& = & 1 -  \alpha({\cal T}) - 
{\epsilon^2 \over 4}\left[ ({\cal T} - 2  M_{rr}) \alpha({\cal T})
+ \int_{\cal T}^\infty  \hspace{-8pt} d{\tau} \;
\alpha({\tau}) \right]  .
\end{eqnarray*}
A bit of rearrangement gives us the weak-signal detection probability
$\gamma({\cal T}) = 1-\alpha({\cal T})-\beta({\cal T})$ as a
function of the threshold:
\begin{eqnarray}
\label{e:betaform}
\nonumber
\epsilon^{-2} \gamma({\cal T}) & = & {1 \over 2} ({{\cal T} \over 2} - M_{rr})\alpha({\cal T}) 
+ {1 \over 4} \int_{\cal T}^\infty  \hspace{-8pt} d{\tau} \; \alpha({\tau})\\
\nonumber
& =  & {1 \over 2} ({{\cal T} \over 2} -  M_{rr})\alpha({\cal T}) +
 {1 \over 2} \sum_{r=0}^{2P} c_r \lambda_r  {\rm e}^{-{\cal T}/2\lambda_r}\\
&= & {1 \over 2} \sum_{r=0}^{2P} \left[ {{\cal T} \over 2} -M_{rr}  +
\lambda_r 
\right] c_r {\rm e}^{-{\cal T}/2\lambda_r}.
\end{eqnarray}
These formulae make it clear that $\gamma = 1-\alpha-\beta$ vanishes as ${\cal
  T} \to 0$ and as ${\cal T} \to \infty$.

\begin{table}
\begin{tabular}{ r|l|l|l|l }
                                       & 1-pt              & 3-pt               &  5-pt   & 7-pt    \\
\colrule
Max ${\gamma \over \epsilon^2} = $     &   0.1424          &   0.1465           &  0.1477 & 0.1483 \\
\@ $\cal T = $                         &   1.548           &   1.863            &  1.918  & 1.942  \\
\@ $\alpha = $                         &   0.3679          &   0.3739           &  0.3767 & 0.3775 \\ 
\colrule
Max ${s \over \epsilon^2 \sqrt{n}} = $ &   0.3113          &  0.3188
       & 0.3204  & 0.3211 \\
\@ $\cal T = $                         &   2.467           &  2.773             & 2.821   & 2.840  \\
\@ $\alpha = $                         &   0.2031          &  0.2093            & 0.2121  & 0.2135 \\ 
\end{tabular}
\caption{The maximum detection probability $\gamma$ and significance
$s$ of the optimal $2P+1$-point peak detection tests, for $P=0,1,2$
and $3$. These correspond to the curves of Fig.~\ref{f:fig5}.  The
{\bf top half} of the table lists the maximum value of the detection
probability $\gamma=1-\alpha-\beta$, and the values of the threshold
$\cal T$ and false alarm probability $\alpha$ for which that maximum
is obtained. The {\bf bottom half} of the table lists the maximum
value of the significance $s$, and the values of the threshold $\cal
T$ and false alarm probability $\alpha$ for which that maximum is
obtained.  }
\label{t:tab2}
\end{table}

It is instructive to return briefly to the $P=0$ (one-point) test.
Eqns.~(\ref{e:alphaform}) and (\ref{e:betaform}) give false alarm and
signal detection probabilities:
\begin{eqnarray*}
\alpha_1({\cal T}) & = &{\rm e}^{-{\cal T}/2 \lambda_0}, \text{ and}\\
& & \\
\gamma_1({\cal T}) &=& 1-\alpha_1 -\beta_1\\
& =&  
  {\epsilon^2 \over 4} {\cal T} {\rm e}^{-{\cal T}/2 \lambda_0}\\
& =& \epsilon^2 \left[- {\lambda_0 \over 2} \alpha_1 \ln
\alpha_1 \right].
\end{eqnarray*}
These should be compared with the resolved-frequency case, given in
Eqns.~(\ref{e:simplealpha}) and (\ref{e:simplegamma}). As expected,
the formulae are identical if $\lambda_0=1$.  However, for the
unresolved frequency case of this Section, Eqn.~(\ref{e:lam0num})
gives $\lambda_0 \approx 0.773695$. Hence the signal detection
probability at a given false alarm probability $\alpha$ is {\it lower}
than in the resolved-frequency case.
\begin{eqnarray*}
\text{For a resolved signal } \gamma&=&-{1 \over 2} \epsilon^2
                                       \alpha \ln \alpha.\\
\text{For an unresolved signal } \gamma&=&-0.3868\; \epsilon^2
                                       \alpha \ln \alpha.
\end{eqnarray*}
Thus, for weak signals, the detection probability of a one-point test
for unresolved signals is 77\% the probability of detection of a
one-point test for resolved signals. 
This can also be
seen by comparing the maxima of the 1-point detection probabilities
shown in Figs.~\ref{f:fig4} and \ref{f:fig5}.

For the first few values of $P$, the detection probability is given by
\begin{eqnarray*}
\epsilon^{-2} \gamma_1({\cal T}) & = &
{\cal T}{e^{-1.38629 - 0.646250\,{\cal T}}} \\
\epsilon^{-2} \gamma_3({\cal T}) & = &
\left ({\cal T}- 0.29663\right ){e^{- 1.17920- 0.638755\,{\cal T}}}\\
& -&\left ({\cal T}-
 1.58708\right ){e^{- 2.85039- 3.63507\,{\cal T}}}\\
&+&\left ({\cal T}- 1.84064\right ){
e^{- 8.12445- 46.4309\,{\cal T}}}\\
\epsilon^{-2} \gamma_5({\cal T}) & = &
\left ({\cal T}- 0.35186\right ){e^{- 1.13581- 0.638380\,{\cal T}}}\\
&-&\left ({\cal T}-
 1.58910\right ){e^{- 2.63902- 3.03752\,{\cal T}}}\\
&+&\left ({\cal T}- 1.89585\right ){
e^{- 8.22249- 44.5006\,{\cal T}}}\\
&-&\left ({\cal T}- 1.91816\right ){e^{- 23.0601-
 6121.0\,{\cal T}}}\\
&+&\left ({\cal T}- 1.91832\right ){e^{- 38.9580- 325142.0\,{\cal T}}}
\end{eqnarray*}
where the subscript on $\gamma$ is $2P+1$: the number of points used
in the test.  Fig.~\ref{f:fig5} shows the detection probability and
significance as a function of false alarm probability $\alpha$ for the
1-, 3-, 5- and 7-point tests, for this case, where the signal
frequency is uniformly distributed in the range $\delta \in \pm 1/2$ a
bin.  It is clear from this Figure, and from Table~\ref{t:tab2} that
while adding the additional information from the nearby frequency bins
does improve the detection probability and significance slightly, the
gain is relatively small. In practice, there is little to be gained
from going beyond the 3- or 5-point tests, as can be seen by noting
that the eigenvalues of ${\bf M}$ drop to small values very quickly with
increasing $P$.  This means that for sensible values of the threshold,
the terms that they add to $\alpha$ and $\beta$ have very small
effects: the dominant terms are from the largest eigenvalues.

\section{INTERPRETATION OF RESULTS AS FREQUENCY SPACE
  ``INTERPOLATION''}
\label{s:freqinterpolate}
In this Section, the optimal statistic $\tau$ of the previous Section
is shown to have a simple intuitive interpretation: it is the total power
contained in a continuous spectrum in the frequency range
$f_{\ell-1/2} < f < f_{\ell+1/2}$.  The continuous spectrum is
obtained from the discrete spectrum $x_j$ via frequency-space
interpolation .

This frequency-space interpolation may be understood in terms of
``zero-padding'', as follows.
\begin{itemize}
\item
Start with the low-resolution frequency-domain Fourier amplitudes
$x_k$ defined by (\ref{e:fftdef}). Here, ``low-resolution'' indicates
that the frequency spacing between successive bins is $1/T$.
\item
Transform these into time-domain $y_j$ for $j=0,\cdots,N-1$.
\item
Zero-pad the time-domain data to $L$ times its original length $N$, by
appending $(L-1)N$ zeros, for $j=N,\cdots,NL-1$.
\item
Now transform back into the frequency-domain to get a
higher-frequency-resolution set of Fourier amplitudes ${\bar x}_k$.
Here ``high-resolution'' indicates that the frequency spacing between
successive bins is $1/LT$.
\end{itemize}
In the limit $L \to \infty$ this gives rise to a continuous
spectrum  ${\bar x}(f)$.  The optimal statistic $\tau$ of the previous
Section is exactly the signal power contained in this continuous spectrum in
the range from $f_{\ell-1/2} < f < f_{\ell+1/2}$. This quantity only depends on the Fourier amplitudes $x_k$ because the zero padding has not added any information to the original data set

To prove this assertion, we first derive a formula for the
high-resolution DFT in terms of the lower-resolution one, following
the procedure above.  The Fourier amplitudes of the time-domain
samples $y_j$ are given by (\ref{e:fftdef}) as
\begin{equation}
  \label{e:fftdef2}
  x_k = \sum_{j=0}^{N-1} y_j \; {\rm e}^{2\pi i j k/N}, \qquad
  \text{for\ } {\scriptstyle k=-N/2+1,\cdots,N/2}.
\end{equation}
The inverse relationship gives the time-domain samples in terms of the
Fourier amplitudes as
\begin{equation}
  \label{e:fftdefinv}
  y_k ={1 \over N} \sum_{j=-N/2+1}^{N/2} \hspace{-8pt} x_j \; {\rm e}^{-2\pi i j k/N}, \qquad
  \text{for\ } {\scriptstyle k=0,\cdots,N-1}.
\end{equation}
Zero-pad these time-domain samples by appending $(L-1)N$ zeros, so
that the total number of time-domain samples is now $NL$. Taking this
back into the frequency domain gives the high-resolution Fourier
amplitudes (for $k=-NL/2+1,\cdots,NL/2$)
\begin{eqnarray}
\nonumber
  {\bar x}_k 
  &=& \sum_{j=0}^{NL-1} y_j \; {\rm e}^{2\pi i j k/NL}  \\
\nonumber
  &=& \sum_{j=0}^{N-1} y_j \; {\rm e}^{2\pi i j k/NL} \\
\nonumber
  &=& {1 \over N} \sum_{j=0}^{N-1} \sum_{r=-N/2+1}^{N/2} 
  \hspace{-12pt} x_r \;
  {\rm e}^{-2\pi i j r/N}{\rm e}^{2\pi i j k/NL}\\
 &=& \sum_{r=-N/2+1}^{N/2} \hspace{-12pt} D_N({k \over L}-r) x_r .
\end{eqnarray}
In the third line, we have carried out the sum over $j$ by using the
geometric series (\ref{e:geometric}).  The last line is the desired
result giving the high-resolution Fourier amplitudes $\bar x$ in terms
of the low-resolution $x$'s. The Dirichlet kernel $D_N$
(\ref{e:DNdef}) is responsible for doing the interpolation.

The high-resolution spectrum has exactly as many degrees of freedom as
the low-resolution spectrum, although it has $L$ times as many
frequency bins. This is because the amplitudes in the high-resolution
spectrum are correlated with each other.  The high-resolution spectrum
also contains an exact duplicate of the low-resolution spectrum.
Since $D_N$ vanishes for non-zero integer arguments, and $D_N(0)=1$,
every $L$'th high-resolution bin contains the same value as one of the
low-resolution bins: $ {\bar x}_{Lr} = x_r$ for all integer $r$.

To finish proving the assertion, we calculate the average power in the
high-resolution frequency bins $k= L(\ell-1/2),\cdots,L(\ell+1/2)-1$.
These $L$ high-resolution bins cover the frequency range from
$f_{\ell-1/2}$ to $f_{\ell+1/2}$, which is $\pm 1/2$ a bin around the
$\ell$'th bin.  Anticipating the final result, this quantity is
denoted ``$\tau$''.  It is
\begin{eqnarray*}
  \tau
  & = &  {1 \over L} \sum_{k=0}^{L-1}
  \left| {\bar x}_{L\ell-L/2+k} \right|^2 \\
  & = & {1 \over L} \sum_{k=0}^{L-1} \left| \sum_{r=-N/2+1}^{N/2}
    \hspace{-8pt} D_N(\ell+{k \over L}-r-{1 \over 2}) \; x_r \right|^2.
\end{eqnarray*}
Since $D_N(x)$ is peaked around $x=0$, in the spirit of the previous
Section, this may be approximated as the sum over the $2P+1$ bins
around the $\ell$'th bin.  Further justification can be found in
Section~\ref{s:window} and in Fig.~\ref{f:fig7}.  This gives
\begin{equation}
\tau  =  {1 \over L} \sum_{k=0}^{L-1} \left| \sum_{r=-P}^{P}
    D_N({k \over L}-{1 \over 2}-r) \; x_{\ell+r} \right|^2.
\end{equation}
In the continuous limit, when the number of high resolution frequency
bins $L \to \infty$, the outer sum can be converted into an integral
over $\delta=k/L-1/2$, giving
\begin{eqnarray*}
  \tau & = & \int_{-1/2}^{1/2} \hspace{-8pt} d\delta \; \; \;
  \left| \sum_{r=-P}^{P} D_N(\delta-r) x_{\ell+r} \right|^2 \\
  & = &  \sum_{r,r'=-P}^{P}  x_{\ell+r} S_{rr'}  x^*_{\ell+r'}.
\end{eqnarray*}
Here, the matrix $S_{rr'}$ is a $2P+1$ dimensional Hermitian matrix
defined by
\begin{equation}
S_{rr'} =  \int_{-1/2}^{1/2} \hspace{-8pt} d\delta \; 
 D_N(\delta-r)  D^*_N(\delta-r').
\end{equation}
This equation should be compared to the definition of $L_{rr'}$ given
in Eqn.~(\ref{e:definem}).  Making the same large $N$
approximation as earlier gives
\begin{eqnarray}
\nonumber
S_{rr'} &=& 
{\rm e}^{i \pi (r-r')\left( 1-{1  \over N }\right)}
\int_{-1/2}^{1/2} \hspace{-12pt} d\delta \; 
 j_0(\pi (\delta-r))  j_0(\pi (\delta-r')) \\
\nonumber
& \approx & 
{\rm e}^{i \pi (r-r')}
\int_{-1/2}^{1/2} \hspace{-12pt} d\delta \; 
 j_0(\pi (\delta+r))  j_0(\pi (\delta+r')) \\
& = & M_{rr'}.
\end{eqnarray}
Thus, the optimal statistic $\tau$ of the previous Section is just the
average power in a continuous interpolated spectrum within a frequency
band of width $\pm 1/2$ a bin around $f_\ell$.

\section{WHY ``WINDOWING'' DOES NOT GIVE A BETTER TEST}
\label{s:window}

Windowing is a well-known method for reducing the bias in a power
spectrum, particularly for frequencies that are not resolved. It is
natural to ask if this technique might provide a better test than the
Neyman-Pearson test.

For large $P$ (the number of bins used on either side of bin $\ell$)
the answer is clearly ``no''.  In this case, the Neyman-Pearson test
{\it is} (by its very definition) the optimal test.  However, if $P$
is very small, one might wonder if windowing could provide a better
test, or if for large $P$, windowing might provide a more efficient
implementation of the optimal Neyman-Pearson test.  The reason is that
in frequency space the amplitudes $|x_k|$ fall off $\propto k^{-1}$
away from the peak.  One might then wonder if windowing can
``concentrate'' more of the power close to the peak, to provide a
better test when $P$ has small values.  As we shall show, the answer
to the question is still ``no'' even when $P$ is small.

``Windowing'' is the process of multiplying the time-domain data $y_j$
by a time-domain window function $w_j$, then transforming the data
into frequency space.  Thus $y_j \to w_j y_j$ in
(\ref{e:fftdef}). This is also referred to as ``apodizing'' or
``tapering''. Note: in addition, one may zero-pad the data set before
taking it into the frequency-domain.  But, as described in
Section~\ref{s:freqinterpolate} the optimal test already effectively
does this, in the limit of infinite zero-padding.

Common choices of windowing functions are given such names as
``Hamming'', ``Parzen'', ``Welch'' and so on.  These window functions
are are chosen for their properties: quickest side-lobe falloff,
narrowest -3db range, minimum spectral bias, and so on.  As an example
here, to explain why windowing the data first does not provide a
better test, we take as a window function the cosine window
\begin{equation}
\label{e:coswindow}
w_j = \sqrt{\frac{2}{3}} \left[ 1 - \cos {2 \pi j \over N} \right].
\end{equation}
The situation for other windowing functions is similar.

\begin{figure}[t]
  \psfig{file=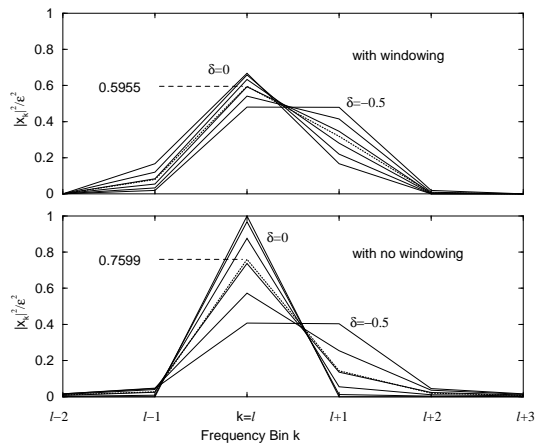,height=7cm,angle=-90}
\caption{The frequency-domain effects of windowing sinusoidal signals
of amplitude $\epsilon$ are shown in the absence of noise. The bottom
graph uses a rectangular window $w_i=1$ (no windowing). The top graph
uses the cosine window defined by Eqn.~(\ref{e:coswindow}). The solid
curves show how the power $|x_k|^2$ is distributed bin-by-bin around
the peak at $k=\ell$, for five different frequencies defined by
$\delta=0,-0.1,\cdots,-0.5$ in
Eqns.~(\ref{e:definefreq}-\ref{e:definefreq2}). The dotted curve shows
the average.  Windowing greatly reduces the difference in $|x_\ell|^2$
between resolved frequencies ($\delta=0$) and unresolved frequencies,
so it reduces the {\it bias} in a spectrum.  However it also reduces
the power in the peak substantially: the mean value is
$0.60\epsilon^2$ with windowing compared to $0.76 \epsilon^2$ without
windowing. This means that windowing does not give a better test: at a
given threshold $\cal T$ it yields a larger false dismissal
probability.}
\label{f:fig7}
\end{figure}

The window function is normalized so that the total power in the
spectrum is the same with or without the window. This is
ensured by the condition (true for large $N$)
\begin{equation}
\sum_{j=0}^{N-1} w_j^2 = N.
\end{equation}
This condition ensure that for stationary noise, the statistical
properties of the noise in the frequency bins is the same with or
without the windowing.  Thus, for example, the expected power spectra
of independent Gaussian-distributed time-domain samples (white
Gaussian noise) are exactly the same for this window and for the
rectangular window $w_j=1$.

Shown in Fig.~\ref{f:fig7} are the spectra of sinusoidal signals
(\ref{e:sigdef}) for the frequency bins near the peak. In the
unwindowed case, a resolved signal ($\delta=0)$ has all its power in
the $\ell$'th bin: $|x_\ell|=\epsilon^2$.  As the frequency shifts
upwards to $\delta=-0.5$, the magnitude of $|x_\ell|^2$ drops to
$0.40\epsilon^2$.  The adjacent ($\ell+1$'th) bin also contains 40\%
of the energy.  The remaining bins contain the other 20\% of the
energy, mostly in bins $\ell-1$ and $\ell+2$.  The large magnitude of
this ratio $1/0.40 = 2.5$ is one reason why rectangular windows are
often undesirable: a peak at a resolved frequency can be as much as a
factor of 2.5 times higher than the same peak at an unresolved
frequency. In contrast, in the windowed case, the magnitude of
$|x_\ell|^2=0.67 \epsilon^2$ when $\delta=0$ and only drops to
$|x_\ell|^2=0.48 \epsilon^2$ when $\delta=-0.5$. The ratio $0.67/0.48
= 1.38$ is much smaller, hence the cosine window produces a less
biased power spectrum than the rectangular window.

But Fig.~\ref{f:fig7} also makes it clear why windowing does not
result in a better test for sinusoidal signals buried in noise than
the Neyman-Pearson test, even for small $P$. The reason is that
windowing ``broadens the peak'' for signals that are near resolved
frequency even more than it ``sharpens the peak'' for signals that are
far from a resolved frequency.  The dotted lines in Fig.~\ref{f:fig7}
show the average power (averaged over the six values
$\delta=0,0.1,\cdots,0.5$. In the windowed case the average power in
the peak is only $0.60 \epsilon^2$ compared to $0.76 \epsilon^2$ for
the unwindowed case.  This reduction in peak power results in a
tremendous loss of significance for small $\epsilon$, when the signals
are buried in noise.  For a given value of the threshold $\cal T$
(corresponding to a fixed false-alarm probability), the windowed signal is
far less likely to cross the threshold when a signal is present than
the non-windowed signal. Thus, it has a higher false dismissal probability
than the Neyman-Pearson test.

Fig.~\ref{f:fig7} also demonstrates that in the unwindowed case,
almost all of the power is within a few bins of the peak.
Consequently even small values of $P$ will give a nearly-optimal test.
For example even for the worst-case signal ($\delta=-0.5$) over 92\%
of the power in contained in just the the range of bins from $\ell-2$
to $\ell+2$.  Averaging over $\delta$, these bins contain more than
96\% of the signal power.  When $P$ is increased this rises rapidly:
in the worse case ($\delta=-0.5$) for $P=10$, the 21 bins around the
peak contain more than 98\% of the total power.  There is effectively
nothing to be gained by increasing $P$ to larger values.

\section{OPTIMAL TESTS IN THE PRESENCE OF NON-GAUSSIAN NOISE}
Section~\ref{s:weaksignal} showed how the weak-signal assumption of
small $\epsilon$ permitted several useful simplifying approximations.
One important simplification was that the optimal statistical test
does not depend upon the amplitude $\epsilon$.

This same weak-signal assumption also makes it possible to find the
optimal statistical test for signals hidden in certain types of non-Gaussian
noise as described, for example, in \cite{upcomingnongauss,upcomingnongauss2}.
Consider the following generalization for the pdf
(\ref{e:restrictedpdf}):
\begin{eqnarray}
\label{e:generalizedpdf}
\lefteqn{p(x|\epsilon)=}  & & \\
\nonumber
& & \int_{-1/2}^{1/2} \hspace{-12pt} d\delta \;
\int_0^{2\pi} {d\phi \over 2\pi}
\prod_{k=-P}^{P} {1 \over 2\pi S_k} \;
{\rm e}^{-g_k \left( { \left|x_{k+\ell} - \epsilon 
 \omega(k+\delta) {\rm e}^{i\phi} \right|^2 \over2 S_k} \right)}.
\end{eqnarray}
The Gaussian case treated in Section~\ref{s:optimalunres} is a special
case of this, for which $g_k(x)=x$ and $S_k=1$.  These types of
non-Gaussian noise models, and the methods that are being used here
(locally optimal tests) are discussed in more detail in
\cite{upcomingnongauss,upcomingnongauss2}, where they are used to construct optimal
search techniques for stochastic background detection and for matched
filtering.

This form of the pdf assumes that the noise in the different frequency
bins is independent, but it allows each bin to have its own,
different, arbitrary statistical distribution.  For example, this can
describe a very common situation, where the pdf has a central Gaussian
region, plus a non-Gaussian tail.  Typically there is a ``knee'' at
some characteristic signal amplitude, where the slope of the
distribution changes, or the non-Gaussian tail begins.  Some
preliminary work \cite{scott99} has shown that it is straightforward
to approximate these functions given a real data stream.

The functions $g_k$ are not completely arbitrary.  In order that
(\ref{e:generalizedpdf}) be properly normalized, one must have
\[
\int_0^\infty du\; {\rm e}^{-g_k(u)} = 1.
\]
For any functional form of $g$, this can be satisfied by adding the
correct constant term to $g$.  We also require that $g$ satisfy the
additional normalization condition
\[
\int_0^\infty du\; u \; {\rm e}^{-g_k(u)} = 1,
\]
which can always be satisfied by re-scaling the argument of $g$.  One
then has
\[
\int dx \; p(x|0) x^*_k x_r  = 2 \delta_{kr} S_k,
\]
so the positive weights $S_k$ can be interpreted as the mean-squared
noise power in the $k$'th frequency bin. This formula should be
compared with Eqn.~(\ref{e:meaninbin}).

\begin{figure}
  \psfig{file=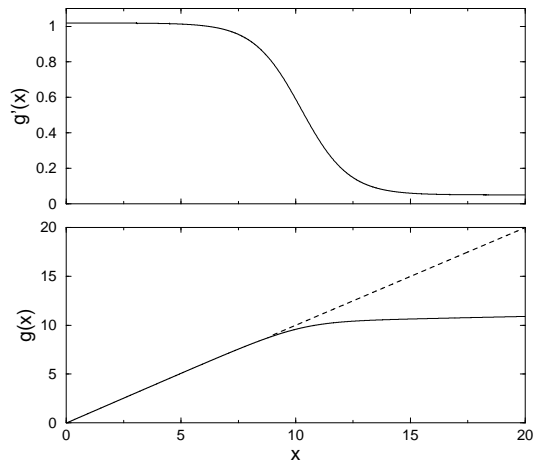,height=7cm,angle=-90} \caption{ An example
  of a function $g(x)$ corresponding to non-Gaussian statistical
  behavior, given by Eqn.~(\ref{e:exampleg}) with $\sigma^2=20$ and
  $p=0.999$. Notice that in the central Gaussian region, $g'(x)\approx
  1$, whereas $g'(x) \to \sigma^{-2}$ when the argument $x$ is larger
  than $\approx \sigma^2/2$.  The dotted line in the bottom graph
  shows (for comparison purposes) $g(x)=x$.  } 
\label{f:fig6}
\end{figure}

For example one might have
\begin{equation}
\label{e:exampleg}
  {\rm e}^{-g(x)} = \kappa 
  \left[ {p } \;  e^{- {\kappa x }}
    +
    {1-p \over \sigma^2} \;  e^{- {\kappa x /\sigma^2}}
  \right],
\end{equation}
where $\kappa = p+ (1-p)\sigma^2$.  Here we assume that $p$ is
positive and less than unity.  The cases of most interest are when
$1-p$ is very small, and $\sigma^2$ is large, so that $\kappa \approx
1$.  Shown in Fig.~\ref{f:fig6} is a graph of $g(x)$ and $g'(x)$ for
the case where $p=0.999$ and $\sigma^2=20$.  This corresponds to a
case where 99.9\% of the data is described by a Gaussian distribution
with unit variance.  The other 0.1\% of the data samples are outlier
points, described by different Gaussian distribution with a variance
of 20.

It is straightforward to derive the optimal peak-detection statistic in
the weak signal limit, by proceeding exactly as in the Gaussian case
of Section~\ref{s:optimalunres}.   We write
\begin{equation}
p(x|\epsilon)=
\int_{-1/2}^{1/2} \hspace{-12pt} d\delta \;
\int_0^{2\pi} {d\phi \over 2\pi} \;
{\rm e}^{W(\epsilon)},
\end{equation}
where $W(\epsilon) =$
\begin{equation}
\sum_{k=-P}^{P} 
\left\{ -g_k \left( { \left|x_{k+\ell} - \epsilon 
 \omega(k+\delta) {\rm e}^{i\phi} \right|^2 \over2 S_k} \right)
-\ln 2\pi S_k \right\}.
\end{equation}
As before, it's easy to see that $p'(x|\epsilon)$ vanishes at
$\epsilon=0$. So the first non-vanishing derivative is
\begin{equation}
  { p''(x|0) \over p(x|0)} = \int_{-1/2}^{1/2} \hspace{-12pt} d\delta
  \int_0^{2\pi}\hspace{-4pt} {d\phi \over 2\pi} \;
  \left[ \left(W'(0) \right)^2 + W''(0) \right].
\end{equation}
The derivatives of $W$ that appear are:
\[
W'(0)=\sum_{k=-P}^{P} {1\over S_k}
g_k' \left( { \left|x_{k+\ell} \right|^2 \over 2 S_k} \right)
 \Re \left( x^*_{k+\ell} \omega(k+\delta) 
{\rm e}^{i\phi} \right)
\]
and
\begin{eqnarray*}
W''(0) &=& - \sum_{k=-P}^{P}
\left\{
  {|\omega(k+\delta) |^2 \over S_k} 
  g_k' \left(
    { \left|
        x_{k+\ell} 
      \right|^2 
      \over 2 S_k} 
  \right) \right.\\
\nopagebreak[4]
  &+& \left. \left[
    {\Re \left(
        x^*_{k+\ell} \omega(k+\delta) {\rm e}^{i\phi}
      \right)
      \over S_k}
  \right]^2  g_k'' 
  \left( { \left|
        x_{k+\ell}
      \right|^2
      \over 2 S_k}
  \right) 
\right\},
\end{eqnarray*}
where $g_k'$ and $g_k''$ are the first and second derivatives of the
function $g_k$ w.r.t. its arguments. Using (\ref{e:intdphi}) to
evaluate the integral over $\phi$, and (\ref{e:definem}) to evaluate
the integral over $\delta$ gives
\begin{eqnarray}
  \nonumber
  \lefteqn{ { p''(x|0) \over p(x|0)} = }&&\\
  \nonumber
  &&  {1 \over 2} \sum_{k,r=-P}^{P}
  { g_k' \left( { \left|x_{k+\ell} \right|^2 \over 2 S_k} \right)
    g_{r}' \left( { \left|x_{{r}+\ell} \right|^2 \over 2 S_{r}} \right)
    \over
    S_{k} S_{r}
    } \;
  x^*_{k+\ell} M_{kr}  x_{r+\ell}\\
\nonumber
&& -
{1 \over 2} \sum_{k=-P}^{P}
  { g_k'' \left( { \left|x_{k+\ell} \right|^2 \over 2 S_k} \right)
    \over
    S^2_{k}
    } \;
  M_{kk}  \left| x_{k+\ell} \right|^2\\
&& -
\sum_{k=-P}^{P}
  { g_k' \left( { \left|x_{k+\ell} \right|^2 \over 2 S_k} \right)
    \over
    S_{k}
    } \;  M_{kk}
\end{eqnarray}
A good algebraic check is to verify that in the absence of a signal
the mean value of this quantity vanishes.

Thus we arrive at the final result: the optimal weak-signal detection
statistic in the non-Gaussian case.  Leaving out the data-independent
constant term, it is
\begin{eqnarray}
  \nonumber
 \tau &= &
  \sum_{k,r=-P}^{P}
  { g_k' \left( { \left|x_{k+\ell} \right|^2 \over 2 S_k} \right)
    g_{r}' \left( { \left|x_{{r}+\ell} \right|^2 \over 2 S_{r}} \right)
    \over
    S_{k} S_{r}
    } \;
  x^*_{k+\ell} M_{kr}  x_{r+\ell}\\
&& -
\sum_{k=-P}^{P}
  { g_k'' \left( { \left|x_{k+\ell} \right|^2 \over 2 S_k} \right)
    \over
    S^2_{k}
    } \;
  M_{kk}  \left| x_{k+\ell} \right|^2.
\end{eqnarray}
This reduces to the original expression (\ref{e:defineopt}) in the
Gaussian case, where $g'=1$ and $g''=0$.  In the non-Gaussian case
(refer to Fig.~\ref{f:fig6}) the effect of the $g'$ and $g''$ terms is
to ``clip'' or ``truncate'' the effects of outlier points.

\acknowledgments 
This research was supported in part by NSF grant PHY-9728704 and
PHY-0071028, and by the Max Planck Society (Albert Einstein Institute,
Potsdam).  We acknowledge useful discussions with S. Frasca,
J. Creighton and E. Flanagan.


\vfill
\end{document}